\documentclass[review]{elsarticle}
\usepackage{hyperref}
\usepackage[T1]{fontenc}
\usepackage{amsmath}
\usepackage[latin1]{inputenc}

\journal{Astroparticle Physics}

\begin{document}

\def\diffnu{\left.\frac{dN_{\nu}}{dE_{\nu}}\right|_{\rm diffuse}}
\def\diffnures{\left.\frac{dN_{\nu}}{dE_{\nu}}\right|_{\rm diffuse}^{\rm res}}
\def\diffnuunres{\left.\frac{dN_{\nu}}{dE_{\nu}}\right|_{\rm diffuse}^{\rm unres}}

\begin{frontmatter}

\title{Prediction of the diffuse neutrino flux from cosmic ray interactions near Supernova Remnants}

\author[bo]{Matthias Mandelartz}

\author[bo]{Julia Becker Tjus\corref{mycorrespondingauthor}}
\ead{julia.tjus@rub.de}

\cortext[mycorrespondingauthor]{Corresponding author}

\address[bo]{Theoretische Physik IV: Plasma-Astroteilchenphysik\\Fakult\"at f\"ur Physik und Astronomie\\Ruhr-Universit\"at Bochum\\44780 Bochum\\Germany}

\begin{abstract}
In this paper, we present high-energy neutrino spectra from 21 Galactic supernova remnants (SNRs), derived from gamma-ray measurements in the GeV-TeV range. We find that only the strongest sources, i.e.\ G40.5-0.5 in the north and Vela Junior in the south could be detected as single point sources by IceCube or KM3NeT, respectively. For the first time, it is also possible to derive a diffuse signal by applying the observed correlation between gamma-ray emission and radio signal. Radio data from 234 supernova remnants listed in Green's catalog are used to show that the total diffuse neutrino flux is approximately a factor of 2.5 higher compared to the sources that are resolved so far. We show that the signal at above 10 TeV energies can actually become comparable to the diffuse neutrino flux component from interactions in the interstellar medium.
Recently, the IceCube collaboration announced the detection of a first
diffuse signal of astrophysical high-energy neutrinos. Directional
information cannot unambiguously reveal the nature of the sources at
this point due to low statistics. A number of events come from close
to the Galactic center and one of the main questions is whether at
least a part of the signal can be of Galactic nature. In this paper,
we show that the diffuse flux from well-resolved SNRs is at least a
factor of 20 below the observed flux. 
\end{abstract}


\end{frontmatter}


\section{Introduction}

The search for the sources of hadronic cosmic rays has made
significant progress within the past few years \cite{becker2008}. In
particular, the detection of high-energy neutrinos and photons has
provided different pieces of information about the origin of cosmic rays:
\begin{itemize}
\item  Gamma-ray emission from the two supernova remnants (SNRs) W44 and IC443 was shown to match the profile of hadronic models \citep{fermi_w44}: The {\it Fermi} satellite has an energy threshold of  $E_{\gamma}\sim 100$~MeV  and is therefore sensitive to the low-energy cutoff from the pion induced gamma-ray spectrum at $\sim 200$~MeV. For W44 and IC443, such a cutoff could be confirmed \cite{fermi_w44}. These sources, on the other hand, show relatively steep spectra towards high energies and cannot accommodate the observed cosmic ray flux up to the knee, i.e.\ up to $E_{\rm CR}=10^{15}$~eV. Today, more than 20 supernova are known to emit at GeV-TeV energies and here, we test the hypothesis that most of the high-energy signal is of hadronic nature.
\item  Astrophysical high-energy neutrinos were detected for the first time with the IceCube detector \citep{icecube2013}. The astrophysical flux persists up to PeV neutrino energies. It is compatible with a power-law spectrum $E^{-\gamma}\cdot \exp\left(-E/E_{\max}\right)$ with an energy cutoff in the PeV range and $\gamma\approx 2$  {\it or} a somewhat flatter spectrum, $\gamma'\approx 2.3$ without a cutoff, $E^{-\gamma'}$ \cite{icecube2014}. The uncertainty on the numbers are relatively large still due to the low statistics and are expected to be improved in the upcoming years by adding more data.
No individual sources have been identified yet, the signal is consistent with an isotropic background. The highest fluctuation is present toward the Galactic center - 8 events cluster towards this direction. It is pointed out by the IceCube collaboration, however, that this excess is not significant at this point.
\end{itemize}
The IceCube detection leaves room for both Galactic and extragalactic
sources. The signal can be explained by extragalactic sources like
galaxy clusters and starburst galaxies/ULIRGs
\cite{he2013,murase2013}, gamma-ray bursts
\cite{cholis2013,winter2013}, extragalactic propagation
\cite{kalashev2013,roulet2013}  and certain types of Active Galactic
Nuclei \cite{stecker2013,becker2014}. Galactic sources have been
discussed in particular in the light of the spatial distribution,
which might hint at a clustering of events close to the Galactic
center. \cite{fox2013} argue that up to two events of the sub-PeV
events can come from TeV unidentified sources in the Galaxy,
i.e.\ sources that lack a radio and X-ray component. Assuming a
correlation between Fermi-detected gamma-rays and high-energy
neutrinos, \cite{neronov2014} predict that the three events that are
fully compatible with coming from the Galactic center region can be
explained by Galactic sources. This investigation of \cite{fox2013}
also includes the two most prominent Milagro sources MGRO J1908+06 and
MGRO J2031+41, which, together with MGRO J2019+37, have been analyzed
in detail in \cite{halzen_kappes2008,gonzalez_garcia2014}. In these
papers, the authors investigate IceCube's capabilities to detect these
in the muon channel, which provides much better directional resolution
compared to the high-energy starting event analysis, where the first
signal was detected. The conclusion is that these individual sources
are difficult to detect within 10 years of lifetime for IceCube. This
result is consistent with the fact that none of the IceCube potential
signal events come from the Cygnus region.  In
\cite{kistler_beacom2006}, the authors calculate the contribution of
Galactic TeV sources to the muon channel, including 14 TeV-detected
sources, with a large fraction of unidentified sources and they
expected a contribution of $1-10$~events per year in IceCube's muon
channel. Today, some of these sources are known better by newer
measurements. For instance, while the emission of RX J1713.7-3946 was
interpreted as hadronic in \cite{kistler_beacom2006} based on the
detection by H.E.S.S., it is now known to most likely be Inverse
Compton emission \cite{FERMI2011-RXJ1713}. This
contribution of leptonic sources reduces the total number of neutrinos
expected. Today's limits to the diffuse gamma-ray emission from the
Galaxy support the hypothesis that the majority of the IceCube signal
should originate from extragalactic sources
\cite{ahlers_murase2014}. Nevertheless, a fraction of the signal could
still be provided by the Galactic component, as it is pointed out in
\cite{razzaque2013,fox2013,padovani_resconi2014,neronov2014}. Thus, in
particular concerning the long-term perspective of high-energy
neutrino telescopes, with KM3NeT being built and both a high- and
low-energy component being discussed for IceCube
\cite{pingu_loi,karle2013}, it is relevant to quantify the possible contribution of Galactic sources to a signal within IceCube.

In this paper, we therefore use those SNRs that have been detected at
gamma-ray energies and that are at the same time well-studied at
multiwavelengths (radio/X-rays). This way, it is possible to both
estimate the contribution from electron synchrotron emission as well
as the hadronic signal at high photon energies.  A correlation
is present between the high-energy gamma-ray flux
$F_{\gamma}(>E_{\gamma})$ and the radio flux at 1~GHz, as already
reported in \cite{fermi_icrc2013}. The existing correlation between the low-
and high-energy signatures is used in this paper to estimate the
expected diffuse neutrino flux using radio data from SNRs identified
at low-energies. We assume that the correlation is linear (see
Fig.\ \ref{radio_gamma:fig}). The scattering of the data points and the size of the
sample still allow for a different type of correlation, but at this
point, assuming linearity seems to be reasonable. Future data are
needed to investigate this behavior in more detail. This is the first time that the contribution from
well-defined supernova remnants in the Galaxy to the diffuse flux is estimated. 

In Section \ref{nu_gamma:sec}, the derivation of the high-energy
neutrino spectra is described, starting with the description of the
multiwavelength-modeling in Section \ref{fit:sec}, followed
by a discussion of the input
parameters (Section \ref{input:sec}), technical details about the
fitting routine (Section \ref{technical:sec})
and the resulting fits in the chosen hadronic scenario in Section
\ref{hadronic:sec}. In Section \ref{nu_spectra:sec}, the resulting
neutrino spectra for the individual remnants are presented and a
diffuse flux is derived. We discuss our results in the context of the
different detection channels and methods of the high-energy neutrino
telescopes IceCube and KM3NeT. In Section \ref{conclusions:sec}, the results are summarized and conclusions are drawn. 

\section{Derivation of individual high-energy neutrino spectra \label{nu_gamma:sec}}
We use multiwavelength data available from 24 SNRs.
The main astrophysical parameters relevant for this
study are listed in Table
\ref{basics:tab}.

\begin{table}
\centering{
\begin{tabular}{c|ccccccc}
\hline\hline
SNR&$d$&$t_{\rm SNR}$&$n_H$&$R_{\rm SNR}$&RA&Dec&Refs\\
&[kpc]&[kyr]&[cm$^{-3}$]&[pc]&&&\\\hline
3C391 & 7.2 & 4.0 & 15.0 & 5.2 & 18h 49m 25s & -00$^{\circ}$ 55' 00" & \cite{Radio1989-Many,Frail1996-3C391Dist,ASCA2001-3C391}\\
W41 & 4.2 & 100.0 & 6.0 & 20.2 & 18h 34m 45s & -08$^{\circ}$ 48' 00" & \cite{Leahy2008-W41}\\
W33 & 4.0 & 1.2 & 6.0 & 1.6 & 18h 13m 37s & -17$^{\circ}$ 49' 00" & \cite{Radio2005-W33,MAGIC2006-W33}\\
W30 & 4.0 & 25.0 & 100.0 & 26.2 & 18h 05m 30s & -21$^{\circ}$ 26' 00" & \cite{Finley1994-W30,Frail1994-W30,FERMI2012-W30}\\
W28 & 1.9 & 33.0 & 140.0 & 13.3 & 18h 00m 30s & -23$^{\circ}$ 26' 00" & \cite{2002AJ....124.2145V,2002AJ....124.2145V,HESS2008-W28}\\
W28C & 1.9 & 0.0 & 100.0 & 2.9 & 17h 58m 56s & -24$^{\circ}$ 03' 49" & \cite{2002AJ....124.2145V,2002AJ....124.2145V,HESS2008-W28}\\
G359.1-0.5 & 7.6 & 5.5 & 1000.0 & 26.5 & 17h 45m 30s & -29$^{\circ}$ 57' 00" & \cite{HESS2008-J1745,1984AAS...57..165R,2002MNRAS.331..537L}\\
G349.7+0.2 & 18.3 & 10.0 & 65.0 & 10.7 & 17h 17m 59s & -37$^{\circ}$ 26' 00" & \cite{2010MNRAS.409..371L,Radio1989-Many,FERMI2010-MULT}\\
CTB 37B & 13.2 & 1.8 & 1.6 & 32.7 & 17h 13m 55s & -38$^{\circ}$ 11' 00" & \cite{Tian2012,MOST-SNRCAT-1996,2009PASJ...61S.197N}\\
CTB 37A & 7.9 & 16.0 & 100.0 & 20.0 & 17h 14m 06s & -38$^{\circ}$ 32' 00" & \cite{MOST-SNRCAT-1996,Tian2012}\\
RX J1713.7-3946 & 3.5 & 1.6 & 0.7 & 30.6 & 17h 13m 50s & -39$^{\circ}$ 45' 00" & \cite{HESS2006-RXJ1713,Cassam2004-RXJ1713,XMM2009-RXJ1713}\\
SN 1006 & 2.2 & 1.0 & 1.0 & 9.2 & 15h 02m 50s & -41$^{\circ}$ 56' 00" & \cite{HESS2010-SN1006,2003ApJ...585..324W}\\
Puppis A & 2.0 & 4.6 & 20.0 & 16.0 & 08h 22m 10s & -43$^{\circ}$ 00' 00" & \cite{1982ApJ...258...22P,2012ApJ...755..141B,Castelletti2006-PupA}\\
Vela Jr & 1.3 & 4.8 & 1.6 & 23.8 & 08h 52m 00s & -46$^{\circ}$ 20' 00" & \cite{Pannuti2010,HESS2007-VelaJr,Combi1999}\\
MSH 11-62 & 6.2 & 1.3 & 7.0 & 11.7 & 11h 11m 54s & -60$^{\circ}$ 38' 00" & \cite{2003ApJ...594..326G,FERMI2012-MSH1162,Radio1986-MSH1162}\\
RCW 86 & 2.3 & 1.8 & 2.0 & 14.1 & 14h 43m 00s & -62$^{\circ}$ 30' 00" & \cite{2003AA...407..249S,Goumard2012-RCW86,2011ApJ...741...96W}\\
W44 & 3.0 & 10.0 & 6.0 & 12.9 & 18h 56m 00s & 01$^{\circ}$ 22' 00" & \cite{Cox1999-W44,Radio1989-Many}\\
G40.5-0.5 & 3.4 & 30.0 & 60.0 & 10.9 & 19h 07m 10s & 06$^{\circ}$ 31' 00" & \cite{2006ChJAA...6..210Y,Green1989MNRAS}\\
W49B & 10.0 & 1.0 & 1000.0 & 4.9 & 19h 11m 08s & 09$^{\circ}$ 06' 00" & \cite{Radio1989-Many,ASCA2005-MULT,FERMI2010-W49B}\\
W51C & 6.0 & 26.0 & 10.0 & 26.2 & 19h 23m 50s & 14$^{\circ}$ 06' 00" & \cite{Moon1994-W51C,ASCA2002-W51C,FERMI2009-W51C}\\
IC443 & 1.5 & 3.0 & 200.0 & 14.2 & 06h 17m 00s & 22$^{\circ}$ 34' 00" & \cite{Reich2003,Welsh2003,Leahy2004-IC443}\\
Cygnus Loop & 0.6 & 15.0 & 5.0 & 25.0 & 20h 51m 00s & 30$^{\circ}$ 40' 00" & \cite{2009ApJ...692..335B,Radio2004-Cygnus,FERMI2011-Cygnus}\\
Cas A & 3.5 & 0.3 & 1.9 & 2.0 & 23h 23m 26s & 58$^{\circ}$ 48' 00" & \cite{Hwang2012-CasA,Reed1995-CasA}\\
Tycho & 3.5 & 0.4 & 0.7 & 4.1 & 00h 25m 18s & 64$^{\circ}$ 09' 00" & \cite{Hwang2002,Suzaku2012-Tycho,CassamChenai2007}\\
\end{tabular}
\caption{Basic parameters of the 24 supernova remnants in this sample: $d$ is the distance to the SNR, $t_{\rm SNR}$ gives the SNR's age, $n_H$ gives the hydrogen density at the interaction site, $R_{\rm SNR}$ represents the size of the remnant and RA/Dec give right ascension and declination. References providing the astronomical data are provided in the last column.\label{basics:tab}}
}
\end{table}

In the model presented here, a one-zone fit is performed to the
observed spectra, taking into consideration leptonic radiation
processes (synchrotron radiation to fit the low-energy peak and
bremsstrahlung as well as Inverse Compton scattering to contribute to
the high-energy peak) as well as hadronic processes ($\pi^{0}$-decays
as a contribution to the high-energy
peak). Similar approaches using
hybrid-emission (leptonic and
hadronic) in a one-zone model
approach have been performed in
e.g.\ \cite{FERMI2009-W51C,halzen2007,berezhko2012,Suzaku2012-Tycho,slane2014,gonzalez_garcia2014}. Previous
approaches focused on the modeling
of one or a few individual
source/sources. Partly, models
included the detailed inclusion of
the acceleration process. We refrain
from doing this, in order to stay as
model-dependent as possible, which
is necessary when considering an
entire population of sources. In
order to estimate the maximum,
diffuse neutrino flux, we will chose
a hadronically dominated scenario
where possible.
In the following,
assumptions to model the spectrum as well as the physical constraints
on the free parameters will be discussed in the context of the
included radiation processes.

\subsection{Modeling of the gamma emission\label{fit:sec}}
The particle number per energy at the source of a species $i=e,\,p$ in
units MeV$^{-1}$ is described as
\begin{equation}
 n_i(E) =
 a_i\left(\frac{\sqrt{E^2+2Em_ic^2}}{\sqrt{E_0^2+2E_0m_ic^2}}\right)^{-\alpha_i}\frac{E+m_ic^2}{\sqrt{E^2+2Em_ic^2}}\tanh\left(\frac{E}{E_{\mathrm{min},i}}\right)\exp\left(-\frac{E}{E_{\mathrm{max,}i}}\right)\,.
\label{crs:equ}
\end{equation}
In general, this is a description of a simple power-law in momentum $p$ with a
low-energy and high-energy cutoff. Parameters are the normalization
$a_i$, the rest mass of the particle, $m_i$, a normalization energy,
chosen to be $E_0=1$~TeV, the minimum energy $E_{\min,i}$
and the maximum energy $E_{\max,i}$.  The minimum energy was chosen to be
$E_{\min,p}=10$~MeV for protons as lower-energy signatures are highly
influenced by ionization
\cite{padovani2009,becker2011,schuppan2012,schuppan2014}. The energy $E$ is the kinetic energy of the
  particle throughout the paper and enters the equation by exchanging
  the differential power-law spectrum in momentum space $dN/dp$ to a
  differential spectrum in kinetic energy, $n$, using the correlation
  $p^2 = \gamma^2 \beta^2 m^2 c^2 = (\gamma^2-1)m^2 c^2$
and the kinetic energy $E=(\gamma-1)m\cdot c^2$.

The
description of synchrotron emission used here is only valid for
$E>m_e\cdot c^2$, which is why the electron's rest mass was used as
the minimum energy. Note that instead of using the typical approach of
a Heaviside function for the low-energy cutoff, we use a hyperbolic
tangent function in order to have a smoother transition.
 The cutoff becomes relevant for the description
of electrons, while it is generally of too low energy to be relevant
for proton-proton interactions. For completeness, we use a
realistic value here, but the cutoff is not crucial for the
description of protons. The high-energy cutoff for protons, $E_{\max,p}$, is kept
variable for those sources that show a cutoff in the gamma-ray data,
while it is fixed at $E_{\max,p}=1$~PeV for those cases with a pure
power-law behavior at high
energies. Concerning the electron
spectra, the high-energy cutoff
$E_{\max,e}$ is fitted, as it in many cases can be
determined by the cutoff of the
synchrotron spectrum at X-ray
energies. 

Based on the particle spectrum at
the source, emitted photons can be
expected from interactions with
surrounding matter. Both leptonic
and hadronic processes have been
taken into account, including in
particular synchrotron radiation of
electrons at low-energies as well as
Inverse Compton scattering, electron
bremsstrahlung and proton-proton
interactions with photon emission at
high-energies. 

\subsubsection{Hadronic gamma-ray
  emission}

The calculation of the hadronic gamma radiation in this work follows
the work of \cite{kelner_pp2006}. The particles emitted by inelastic
proton-proton scattering are given by the following formula, up to
energies of $E<100$~GeV:
\begin{equation}
 \frac{\mathrm dn}{\mathrm dE} = \tilde{n}\int\limits_{E_\mathrm{min}}^\infty cn_\mathrm{H}\frac{2}{\kappa\sqrt{E_\pi^2-m_\pi^2c^4}} \sigma_\mathrm{inel}\left(m_\mathrm{p}c^2 + \frac{E_\pi}{\kappa}\right)\frac{n_\mathrm{p}\left(m_\mathrm{p}c^2 + \frac{E_\pi}{\kappa}\right)}{4\pi\,d^{2}}\,\mathrm dE_\pi\,,
\label{delta_approx:equ}
\end{equation}
where $\tilde{n}$ is used for the continuity of the function at higher
energies, $\kappa$ the fraction of the kinetic energy of protons that
is released in secondaries during the decay of $\pi$ and $\eta$
mesons, and $n_\mathrm{H}$ the number density of the interacting
gas. $\sigma_{\rm inel}$ is the total cross section.

Above $100\,$GeV the production rate approximation, usually called the
delta-approximation, can be improved by taking into account the
distribution function $F_i$ that was modeled upon the numerical data
provided by the SIBYLL code (see \cite{kelner_pp2006} for details). For the
differential flux density for a secondary particle species $i=e,\,\mu,\,\nu,\,\gamma$ with an energy above $100\,$GeV
\begin{equation}
 \frac{\mathrm dn}{\mathrm dE_i} = \int\limits_{E_i}^\infty c n_\mathrm{H} E_\mathrm{p}^{-1}\sigma_\mathrm{inel}\left(E_\mathrm{p}\right) n_\mathrm{p}\left(E_\mathrm{p}\right)F_i\left(\frac{E_i}{E_\mathrm{p}},E_\mathrm{p}\right)\,\mathrm dE_\mathrm{p}
\label{kelner:equ}
\end{equation}
is obtained. It is important to note here that this formalism includes
the secondaries produced by $\eta$ mesons additionally to the ones
generated by $\pi$ mesons.

The neutrino spectra are automatically determined by the hadronic fit
of the gamma-ray spectra. The total neutrino spectrum at the source is
of about the same order as the gamma-ray spectrum. In our graphs, we
only show muon- and anti-muon neutrinos, which is why we divide the
total neutrino flux, concretely calculated via the formalism described
in \cite{kelner_pp2006}, by a factor of 3, taking into account oscillations.

\subsubsection{Synchrotron
  Radiation}
According to \cite{BlumenthalGould} the in synchrotron radiation power
emitted by a population of electrons with a number of particles per
energy interval $n_e$ can be described by \cite{CrusiusSchlickeiser1986}
\begin{equation}
 \frac{\mathrm dn}{\mathrm d\epsilon} =\int\limits_{mc^2}^\infty \frac{\sqrt{3}\pi e^3B}{2h\epsilon mc^2}n_e(E'-mc^2)\frac{\epsilon\sin\alpha}{\epsilon_c} \mathrm{CS}\left(\frac{\epsilon\sin\alpha}{\epsilon_c}\right)\,\mathrm dE'\,,
\end{equation}
where $B$ is the magnetic field strength, $\alpha$ is the angle
between particle velocity and magnetic field, $n_e$ is the
differential electron density, and $\epsilon_c$ is the critical energy known as
\begin{equation}
 \epsilon_c = \frac{3heB}{2 m c}\frac{E'^2}{m^2c^4}\sin\alpha\,.
\end{equation}
In the above equations, $E'$ is the total energy of the particle, $E'=E+m\cdot c^2$.
The CS-function is here defined as a product of Whittaker-functions, as follows:
\begin{eqnarray}
 \mathrm{CS}(x) &=& W_{0,\frac{4}{3}}(x)W_{0,\frac{1}{3}}(x)-W_{\frac{1}{2},\frac{5}{6}}(x)W_{-\frac{1}{2},\frac{5}{6}}(x)\,.
\end{eqnarray}
The flux at Earth is achieved by taking into account the decrease of
the signal with $4\pi d^2$.

\subsubsection{Bremsstrahlung}
The Bremsstrahlung emission in this work is calculated following the
work of \cite{BlumenthalGould}, its differential cross-section is given as
\begin{equation}
 \frac{\mathrm d\sigma}{\mathrm d\epsilon} = \frac{\alpha_f r_0^2}{\epsilon E_i^2} \left[(E_i^2+E_f^2)\phi_1-\frac{2}{3}E_iE_f\phi_2\right]\,,
\end{equation}
which describes a particle with energy $E_i$ decelerating to the
energy $E_f$ and producing a photon of the energy $\epsilon
=E_i-E_f$. Here, $\alpha_f$ is the fine-structure constant and $r_0$
is the classical electron radius. The functions $\phi_j$ with $j=1,\,2$ depend on the nature of the particle's Coulomb field the incident particles scatter with. In case of an unshielded particle with the charge $Ze$, e.g. electron or proton for $Z=1$, the functions become $\phi_1=\phi_2=Z^2\phi_u$ with 
\begin{equation}
 \phi_u = 4\left(\ln\frac{2 E_i E_f}{\epsilon mc^2}-\frac{1}{2}\right)\,.
\end{equation}
In case of a nucleus-one-electron system the functions $\phi_j$ start to be more complicated to evaluate. \hspace{0mm}\cite{BlumenthalGould} state that their shape in this case is
\begin{equation}
 \phi_j = (Z-1)^2\phi_u+8Z\left(\alpha_j+\int\limits_\delta^1 f_j(q)\left[1-\left(1+\frac{q^2}{4\alpha_{f}^{2} Z^2}\right)^{-2}\right]\,\mathrm dq\right)\,,
\end{equation}
where $\alpha_1 = 1$, $\alpha_2 =
5/6$, and $\delta=\epsilon
mc^2/E_iE_f$. The simple functions
$f_j(q)$, which also depend on
$\delta$, are given by \cite{BlumenthalGould} as:
\begin{eqnarray}
 f_1 &=& \frac{(q-\delta)^2}{q^3}\,,\nonumber\\
 f_2 &=& \frac{q^3-6\delta^2q\ln\frac{q}{\delta}+3\delta^2q-4\delta^3}{q^4}\,.
\end{eqnarray}
It also needs to be taken into
account that molecular hydrogen has
the highest abundance in galactic
molecular clouds, while only
one-electron nuclei and unshielded
ones can be dealt with. However,
molecular hydrogen can be
approximated as two hydrogen atoms,
with an error below 3\%
\citep{Gould1969,Schlickeiser2002}. The
  molecular content of the interaction environment is subject of the
  concrete object under investigation. Some of the remnants, like
  CasA, appear to live in the thin, interstellar medium, while others
  like W44 and W51C are embedded into molecular clouds.

The resulting total spectrum for the bremsstrahlung generated photons at the source can be expressed readily by the sum over the particle species $s$ as
\begin{equation}
 \frac{\mathrm dn}{\mathrm d\epsilon} = c\int\limits_{\epsilon+mc^2}^\infty n_e(E'-mc^2)\sum\limits_s n_s\frac{\mathrm d\sigma_s}{\mathrm d\epsilon}\,\mathrm dE'\,.
\end{equation}
To include helium as well as a scattering partner for the electrons, a factor of 1.3 is used as suggested by \cite{Schlickeiser2002} for the interstellar medium as follows:
\begin{equation}
 n_{{\mathrm{H}_\mathrm{I}}}\frac{\mathrm d\sigma_{\mathrm{H}_\mathrm{I}}}{\mathrm d\epsilon}+n_{\mathrm{H}_2}\frac{\mathrm d\sigma_{\mathrm{H}_2}}{\mathrm d\epsilon}+n_\mathrm{He}\frac{\mathrm d\sigma_\mathrm{He}}{\mathrm d\epsilon}\approx 1.3 \left(n_{{\mathrm{H}_\mathrm{I}}}+2n_{\mathrm{H}_2}\right)\frac{\mathrm d\sigma_{{\mathrm{H}_\mathrm{I}}}}{\mathrm d\epsilon}\,.
\end{equation}
Finally, to receive the spectrum at Earth, the result needs to be divided by $4\pi\,d^2$.
\subsubsection{Inverse Compton
  Scattering}
\cite{Jones1968} derived the following result for the Compton spectrum of a single electron:
\begin{equation}
 \frac{\mathrm dN_\gamma}{\mathrm d\epsilon'\mathrm d\epsilon} = cn(\epsilon')\frac{2\pi r_0^2 m^2c^4}{E'^2}\frac{1}{\epsilon'}\left[2q\ln q+(1+2q)(1-q)+\frac{1}{2}\frac{(\Gamma q)^2}{1+\Gamma q}(1-q)\right]\,,
\end{equation}
where $r_0$ is the classical electron radius, $\epsilon'$ is the
energy of the photon before scattering, $n(\epsilon')$ is the differential photon density, and the two dimension-less parameters $\Gamma$ and $q$ are defined as
\begin{equation}
 \Gamma = \frac{4\epsilon'E'}{m^2c^4}, \quad q = \frac{\epsilon}{\Gamma_\mathrm{e}\left(E'-\epsilon\right)}\,.
\end{equation}
The photon density is obtained by adding the different sources of photons in that specific region. The Inverse Compton radiation was calculated for scattering on the CMB only, IR emission from dust and stellar radiation fields might contribute to the emission \cite{Suzaku2012-Tycho}, but are not generally known for every source.\\
To obtain the spectrum emitted by the source a convolution of the the single electron spectrum with the differential spectrum of applicable electrons has to be considered:
\begin{equation}
 \frac{\mathrm dN}{\mathrm d\epsilon} =
 \int\limits_0^\infty\int\limits_{E_\mathrm{min}}^{\infty}
 n_e(E'-m\cdot c^2)\frac{\mathrm dN_\gamma}{\mathrm d\epsilon'\mathrm d\epsilon}(E',\epsilon',\epsilon)\,\mathrm dE'\mathrm d\epsilon'\,,
\end{equation}
where the minimum energy $E_\mathrm{min}$ results from the kinematics of the problem \citep{BlumenthalGould}, leading to the inequality $\epsilon(1+\Gamma)\leq \Gamma E'$ and can be expressed as
\begin{equation}
  E_\mathrm{min} = \frac{\epsilon}{2}\left[1+\sqrt{1+\frac{m^2c^4}{\epsilon\epsilon'}}\right]\,.
\end{equation}
Again, the flux at Earth is given by dividing the result by $4\pi d^2$.
\subsubsection{Secondaries and
  Densities}
The density for the target
  material for bremsstrahlung and proton-proton interactions was fixed
  to the same amount taken from literature about the specific
  remnants (see Table \ref{basics:tab}). In this approach, we neglect any contribution from
  secondary electrons and positrons which are produced via the decay
  of charged pions, the latter being co-produced with the neutral
  pions. While these processes are being discussed as a possible
  dominant source of electrons in starburst galaxies (see
  e.g.\ \cite{loeb_waxman2006,lacki_beck2013}), they can be
  neglected in the much less dense environment of SNRs  in the Milky Way. The main reason is the much lower densities in this
    region which lead to optical depths for proton-proton interactions
    much smaller than one: The optical depth is given as $\tau_{\rm pp}=R\cdot n_{\rm H}\cdot
  \sigma_{\rm inel}$. For an order-of-magnitude estimate, we can use a
  fixed value for the size of the interaction region, $R\sim 10$~pc
  and the proton-proton cross section, $\sigma_{\rm inel}\approx 3\cdot
  10^{-26}$~cm$^2$ \citep[e.g.]{bb2009}. With those values, the optical
  depth is $\tau_{\rm pp}\approx 10^{-6} \cdot (n_{\rm H}/{\rm 1
    cm}^{-3})$. Given the fact that the spectrum of secondary
  electrons is approximately proportional to the primary proton
  spectrum multiplied by the optical depth as well as a factor $<1$ as
  the fraction of energy going from the initial proton into the final electron, we have a ratio
  between protons and secondary electrons of $K_{\rm ep,sec}<
   10^{-6} \cdot \left(n_{\rm H}/{\rm 1
    cm}^{-3}\right)$. This ratio is much larger than
  what we expect from primary electrons (see below) and thus, the 
  contribution from secondaries can be neglected here. It is not a
  contradiction that the optical depth is low. Although the production
  efficiency is low, the number of cosmic rays is sufficient to
  produce a detectable gamma-ray signature. There turn out to be five SNRs in our sample where the production rate of secondary electrons and positrons actually does resemble the observed synchrotron luminosity. We investigate the question whether or not it is reasonable to neglect secondaries for the production of synchrotron radiation for these specific sources in detail in Section \ref{secondaries_result_discussion:sec}.

The scenario is different for
  starburst galaxies, where the size of the interaction region is
  significantly larger and the average density is typically much
  larger as well. Due to higher magnetic fields, the synchrotron loss time scale is generally shorter as well, so that a larger fraction of electron energy goes into synchrotron production. Note that the observed
    gamma-ray emission from
    the starburst galaxies M82 and NGC253 otherwise is likely to
    resembel the diffuse emission comparable to what is calculated
    here: the observed gamma-ray spectra are much flatter than the
    dominant gamma-ray emission in our Galaxy. This indicates that the
    interaction must happen rather close to the remnant, where
    diffusion has not steepened the spectra yet. An alternative
    explanation would be very different primary cosmic ray spectra,
    but given the much higher density in starburst galaxies, it is
    quite realistic to have dominant interactions close to the sources.
\subsection{Free parameters and physical restrictions\label{input:sec}}
In the initial modeling procedure,
the following parameters are
considered to be free. Note that
this approach is firstly only done
for the fitting procedure, and the
physically relevant cases are
selected afterwards as described in
the introduction. In this
subsection, we discuss, in which range they have to lie concerning astrophysical constraints. Scenarios
which do not lie in the physically relevant range are discarded.
\paragraph{Magnetic field}
The modeling is done for a fixed range of magnetic fields to begin
with, starting from the best-fit magnetic field when only considering
leptonic processes, $B=B_{\rm inf}$, the magnetic field is moved to
larger values successively, including hadronic processes as
well. Here, the maximum tested value is $B=5B_\mathrm{inf}$. This range is compatible from what is
usually observed in SNRs, see e.g.\ \citep{uchiyama2007}. There is one
exception which is W44, which we simulate over a wider range in order
to reach the case of hadronically dominated contribution. For a second
source, W28, the simulated range also turns out to be non-efficient:
in this one-zone model, magnetic fields of larger than
$500$~$\mu$Gauss are needed in order to reduce the Inverse Compton
scattering to below the measured data. These magnetic field values
seem unrealistic and we therefore model W28 in a two-zone approach,
where synchrotron radiation happens in a different environment than
the hadronic interactions. We use a magnetic field strength of
$11$~$\mu$Gauss for our final fit.

\paragraph{Total energy budget}
The total energy derived for a particle density per energy interval, $n_i$, with particles of the species $i$ is calculated as
\begin{equation}
 E_{i, {\rm tot}} = \int\limits_{0}^\infty n_i(E)\cdot E\,\mathrm dE\,.
\label{etoti:equ}
\end{equation}

The total, non-thermal energy is then given as the sum of energy put into electrons, protons and the magnetic field, 
\begin{equation}
E_{\rm tot}=E_{\rm p,tot}+E_{\rm e,tot}+E_{\rm B, tot}
\label{etot:equ}
\end{equation}
with $E_{\rm B,tot}=u_{\rm B}\cdot V=B^{2}/(2\,\mu_0)\cdot V$ in SI-units, or $E_{\rm B, tot}= B^{2}/(4\,\pi)\cdot V$ in cgs-units.
We assume that the density is homogeneously distributed over the
entire remnant, i.e.\ $V = 4/3\cdot \pi \cdot R_{\rm SNR}^{3}$.  Note that the total energy budget of protons and
  electrons does not reflect the gamma-ray budget for hadronic or
  leptonic emission, as optical depths for the different processes
  differ from each other. 

The total
non-thermal energy of the SNR is not allowed to
exceed $10^{51}$~erg to be consistent with the maximum energy available
from the kinetic energy put into the
SNR, considering a star of a few
solar masses. Two exceptions, W51C ($\sim 2.3\cdot 10^{51}$~erg) and W30 ($\sim 1.1\cdot 10^{51}$~erg) lie slightly above this value, but are compatible with a non-thermal energy budget of $<10^{51}$~erg within the errors, especially considering that the target density $n_H$ enters linearly and the radius of the emission region $R_{\rm SNR}$ enters cubed in the calculation. The latter in particular is considered to rather provide an upper limit to the non-thermal energy budget of an SNR, as it is expected that the different components (electrons, protons and magnetic field) are not necessarily expected to  fill the entire remnant. 
\paragraph{Spectral indices of primary spectra}
The power-law spectra of the primary electrons and protons are
expected to be produced via diffusive shock acceleration \citep{fermi1949,fermi1954,krymskii1977,bell1978a,schlickeiser1989a,schlickeiser1989b}. While in extremely relativistic environments, spectra
significantly flatter than $\alpha_i\sim 2$ can be achieved \citep[e.g.]{stecker2007,mbq2008}, in an SNR environment, the spectra
are believed to be  $\alpha_i\sim 2$
or steeper ($i=e,\,p$). Thus, only spectra
which are compatible with $\alpha_i> 2$ within the statistical
errors are taken into account.
\paragraph{Ratio of electron-to-proton energy}
For primary spectra with a spectral index of $\alpha_i=2.2$, the ratio
of the energy in protons to electrons is expected to lie around
$K_{\rm ep}=E_{\rm e}/E_{\rm p}\approx 0.01$ \citep{Schlickeiser2002}. This ratio is strongly
dependent on the integration limits for the both particle species and
on the spectral indices of the primary spectra
\citep{merten2014}. For extragalactic cosmic
rays, for instance, observations indicate that the ratio is rather
$K_{\rm ep, extrag}\approx 0.1$ or even larger \citep{wb1997,Winter2014}. Depending on the
maximum energies and the spectral indices of the two processes,
theoretically received results even show that the ratio can become
larger than 1. While the ratio is usually
  discussed to be $K_{\rm ep}\ll 1$ (see e.g.\ \cite{Schlickeiser2002,kang2014}), these areguments are based on the
  assumption that electrons and protons have the same spectral
  behavior. This is not the case for the remnants considered here. If
  the dependence on the spectral index is considered, it is not enough
  to consider the ratio of differential particle spectra. Here, the
  total energy budget needs to be considered instead. For standard
  values of $\alpha_e=\alpha_p\approx 2.2$, following the calculation
  of \cite{Schlickeiser2002}, we receive a ratio close to
  $K_{ep}\approx 0.01$ as discussed in the literature. Deviating from
  these standard values, the ratio can vary significantly. As our
  spectral indices deviate quite significantly from $2.2$ and in
  particular are not the same for electrons and protons in the same
  remnant, the standard scenario does not apply and values much
  smaller than 0.01 but also larger than this value can be
  acheived. On average, the ratio of electrons to protons can still be
  0.01 as observed at Earth, but for each individual remnant, $K_{rm
    ep}$ can deviate quite significantly from the standard value. The
  problem of different indices for protons and electrons also implies
  an additional uncertainty from integration: uncertainties in the
  integration limits are certainly present. These facts make
  it necessary to chose a rather large range of values. For all remnants, we allow the
ratio to vary between $10^{-4}<K_{\rm ep}<50$. There are
two remnants, MSH 11-62 and W44, with a ratio larger
than one (in the case of MSH 11-62, $K_{ep}\sim 10$ and for W44,
$K_{ep}\sim 50$). For the given densities, it is difficult to reduce
the values even further and we believe these high values are an
artifact of poorly known integration limits. In particular, for W44,
the Fermi collaboration could show that the spectrum of W44 is likely
to be of hadronic nature \cite{fermi_w44}, which is a confirmation for us to include
this source.
Scenarios outside this range are
discarded. With respect to the
the typically used values of $0.01-0.1$ in the literature, this may
appear as a rather large range. However, there is no fully developed argument from
theory that fixes the ratio between electrons and protons to a certain
value and even with the usually cited standard calculation, the
result strongly depends on the input parameters. This is why we keep
this ratio within this larger range. Detailed results on the
theoretical calculation of the variation of the parameter $K_{\rm ep}$
are in preparation \citep{merten2014}. 

\subsection{Modeling procedure\label{technical:sec}}
In this paper, we use the following scheme to systematically model the
spectra: 
\begin{enumerate}
\item We include all above mentioned radiation processes into the
  fitting procedure
\item In order to minimize $\chi^2$ for the fit, the Nelder-Mead \citep{NelderMead}
  algorithm from the GSL \citep{GSL} is used.
\item The leptonic and hadronic parts are fit separately. We start with a purely leptonic in order to
  determine a minimum magnetic field necessary in order to describe
  the spectral energy distribution. Increasing the magnetic field from this minimum value
  will give more weight to the hadrons: as the synchrotron emission is
  fixed by data, less electrons are needed to produce the
  synchrotron bump when increasing the
  magnetic field. Thus, contributions from Inverse Compton emission
  and bremsstrahlung decrease with the magnetic field strength and
  the lacking energy in the high-energy part of the spectrum is
  provided by hadrons. The magnetic field range
  investigated here is within a factor of five of the originally
  determined minimal value. In this procedure, the gas density is
  fixed as indicated in Table \ref{basics:tab}. For each fit, the
  magnetic field is fix as well, and we receive a set of best-fit
  options in the magnetic field range we chose.
\item Finally, out of these best fit scenarios, we chose a scenario
  that is, if possible, hadronically dominated and at the same time
  obeys the boundary conditions defined in Section
  \ref{input:sec}. In most cases, this means to
    increase the magnetic field up to a value where the hadronic
    gamma-ray emission dominates the spectral energy
    distribution. Going beyond this value often implies to have
    extremely large magnetic fields, connected to an extreme total,
    non-thermal energy budget. There is a small number of exceptions
    from this general rule,
    each of which is discussed below.
\end{enumerate}

\begin{figure}[ht]%
 \begin{center}%
 \includegraphics[width=0.9\linewidth]{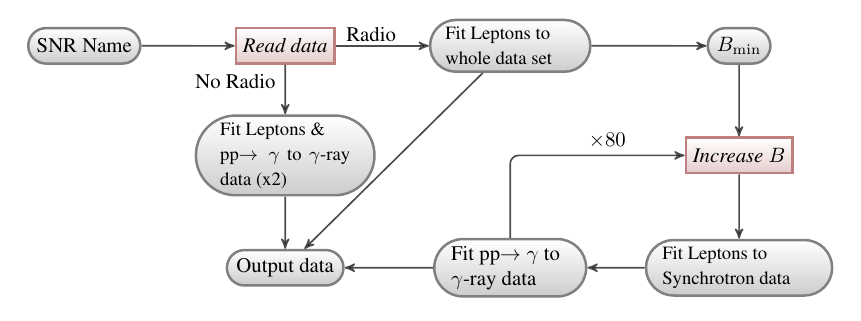}
 \end{center}%
 \caption{Scheme of the work flow of the fitting routine. \label{scheme:fig}}
\end{figure}%

\subsection{Hadronic scenario\label{hadronic:sec}}
According to the fitting procedure described above, we now chose a scenario for which hadrons play a dominant role in the high-energy emission process - if possible within the parameter range discussed above. In general, raising the magnetic field suppresses the leptonic signature and at some value of $B$, the high-energy part is usually dominantly described by hadrons. Going towards higher magnetic fields does not change this and in accordance with keeping the total non-thermal energy below $10^{51}$~erg, we typically chose a fit at the lowest magnetic field possible.

A hadronically-dominated fit turns out to be possible in 21 out of the 24 cases. Three cases turn have to be fit leptonically: 
\begin{enumerate}
\item {\bf RX J1713.7-3946} shows an extremely flat spectrum with a
  cutoff at TeV energies \cite{FERMI2011-RXJ1713}, matching an Inverse Compton scenario. The spectrum is as flat as $\sim E^{-1}$, which does not fit the typical diffusive shock acceleration scenario, see e.g.\ \cite{bell1978a,bell1978b,schlickeiser1989a,schlickeiser1989b}.
\item {\bf RCW 86} also reveals an Inverse Compton-like shape with a
  very flat spectrum up to a cutoff at TeV energies and does not match
  a hadronic scenario \cite{fermi_rcw2014}.
\item {\bf G359.1-0.5} is not compatible with the cutoff in the
  $\pi^{0}-$spectrum at a few hundred MeV, but continues as a
  power-law toward lower energies \cite{FERMI2011-J1745}.
\end{enumerate}
For the remaining 21 sources,
Figures \ref{etot0:fig},
\ref{etot1:fig}, \ref{etot2:fig},
\ref{etot3:fig}, \ref{etot4:fig} and \ref{etot_sn1006:fig} in the appendix
show the non-thermal
energy budget needed for each source
at the B-field values tested
here. The horizontal line indicate
the concrete values we chose here in
order to have hadronically dominated
emission: in order to minimize the
energy, we always take the lowest
magnetic field at which hadrons
start to dominate the emission
process. The fit parameters for this
specific scenario are listed in
Tables \ref{electrons:tab} (electron
parameters) and \ref{hadrons:tab}
(proton parameters). The spectral energy distribution (SED) fits
are also shown in the appendix, see
Figures \ref{sed0:fig},
\ref{sed1:fig}, \ref{sed2:fig},
\ref{sed3:fig}, \ref{sed4:fig} and \ref{sed_sn1006:fig}.

\begin{figure}[ht]%
 \begin{center}%
 \includegraphics[width=0.9\linewidth]{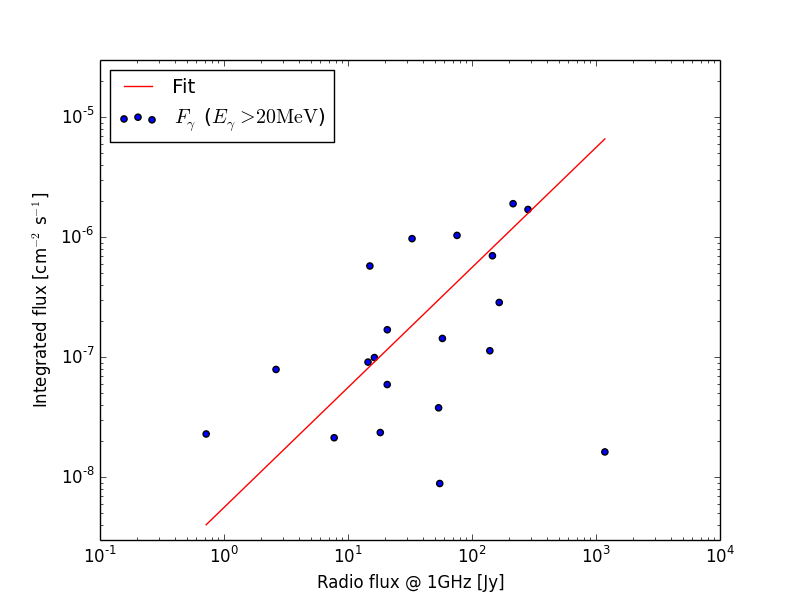}
 \end{center}%
 \caption{Integrated gamma-ray flux,
   $F_{\gamma}(E_{\gamma}>20$~MeV$)$, versus radio flux at 1 GHz,
   $F_{\rm radio}(\nu=1$~GHz$)$ for 21 SNRs with a potentially
   hadronic high-energy signature. The correlation between the two
   quantities is compatible with a linear one, $F_{\gamma}=
   A_{\gamma\,\rm radio}\cdot \left(F_{\rm radio}/{\rm Jy}\right)$,
   with $A_{\gamma\,\rm radio}=5.59882298\cdot
   10^{-9}$~cm$^{-2}$~s$^{-1}$. Here, only CasA, which is
   exceptionally bright in the radio, has been excluded from the
   fit. It is clear that this correlation also
     could deviate from linearity, given that the scatter is
     relatively large. The tendency of a correlation is still clear and as a working
     hypothesis, we use a linear correlation. In the future, with
     larger statistics, the question of linearity can be investigated
     in more detail.   \label{radio_gamma:fig}}
\end{figure}

The multiwavelength fits can now be
used to investigate the correlation
between the low-energy signature
arising from electron synchrotron
radiation and high-energy photon
signal. In
Fig.\ \ref{radio_gamma:fig}, the
total gamma-emission, integrated
from 20 MeV, versus the radio flux
at one GHz is shown. A correlation
is present between the two energy
bands, which is also seen when only
considering Fermi data, see
\cite{fermi_icrc2013}. As the
gamma-emission is fully dominated by
hadronic processes in our
calculation, we find that the
hadronic component can be assumed to
scale directly with the radio flux,
$F_{\rm had}\propto F_{\rm
  radio}$. This provides us with a
method to estimate the signal to be
expected from SNRs that do not show
a signal in gamma-rays yet, i.e.\ so far unresolved hadronic sources. In Green's catalog \cite{green_cat}, 274 SNRs are listed, with most of them having a measured radio flux around 1 GHz. It will be discussed in Section \ref{diffuse_nus:sec} how we will use this piece of information in order to derive the neutrino flux from SNRs so far unresolved in the high-energy regime.

\begin{figure}[ht]%
 \begin{center}%
 \includegraphics[width=0.9\linewidth]{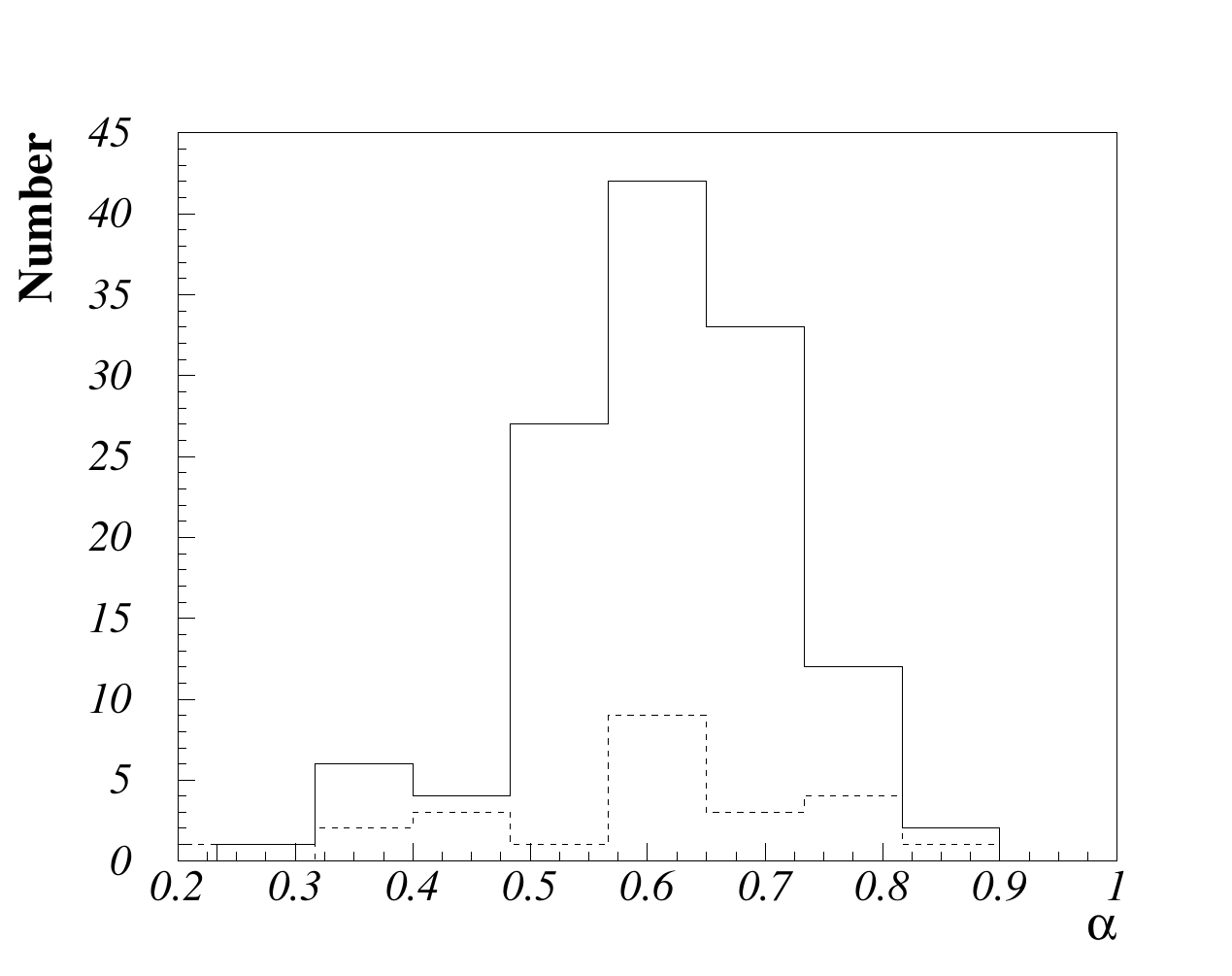}
 \end{center}%
 \caption{Radio
   spectral indices for those 127 SNRs in the Green catalog which have a well-defined index at 1 GHz (solid line). The dashed line shows the distribution for the sub-sample investigated in this paper. Both distributions scatter around $\alpha_{\rm radio}\sim 0.6$, corresponding to $\alpha_e\sim 2.2$.   \label{alpha_radio:fig}}
\end{figure}

\begin{figure}[ht]%
 \begin{center}%
 \includegraphics[width=0.9\linewidth]{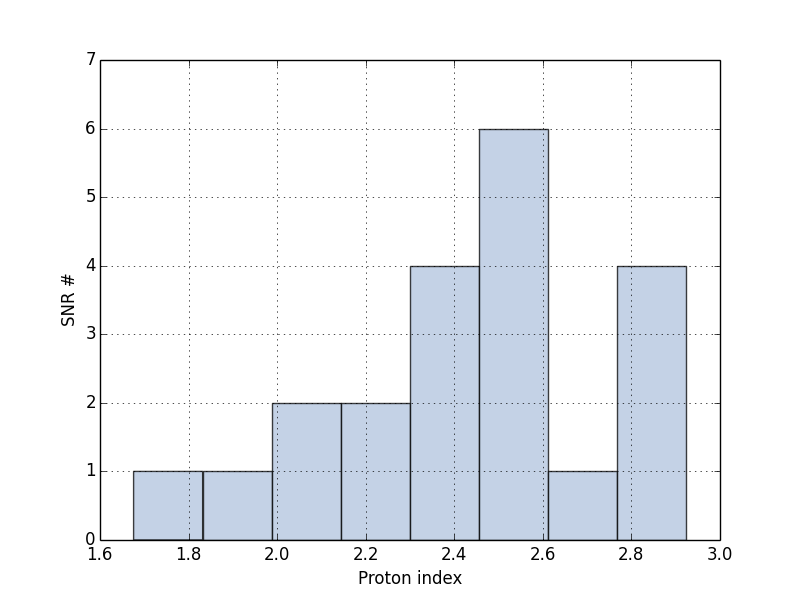}
 \end{center}%
 \caption{Proton spectral index distribution for the 21 SNRs in the sample fitted in this paper. A main peak is present at around $\alpha_p\sim2.4$ and a smaller population has indices around $\alpha_p\sim 2.9$.   \label{proton_index:fig}}
\end{figure}

In Fig.\ \ref{alpha_radio:fig}, we show the radio spectral index for all 127 SNRs in the Green catalog that have a well-defined spectral index in the 1~GHz region, compared to the same index distribution for the sub-sample of sources fitted in this paper. Both distributions peak at around $\alpha_{\rm radio}\sim 0.6$, which translates into an electron index of $\alpha_e = 2\cdot \alpha_{\rm radio}+1$ = 2.2. This is a cross-check that the fits for our sub-sample represent an average sample of SNRs, with the sub-sample being distributed approximately as the larger sample. Figure \ref{proton_index:fig} shows the distribution we find for the protons. Compared to the electrons, the protons show a generally broader distribution. Note that those two sources with indices in the bins below $\alpha_p=2.0$ are compatible within errors with a spectrum as flat as $E^{-2}$, in accordance to the parameter range set in Section \ref{input:sec}. The distribution appears quite inhomogeneous, with a larger population of sources with extremely steep spectra (up to $\alpha_p\sim 2.9$). This part of the population has been discussed to be older SNRs in molecular clouds, while younger remnants in a less dense environment have flatter spectral indices. The question of the steepness of these spectra, also in combination with the high-energy cutoff, is an important question concerning the origin of cosmic rays themselves, which we cannot answer in this paper: the diffuse cosmic ray spectrum as observed at Earth follows a spectrum close to $E^{-2.7}$ up to a cosmic ray energy of $1$~PeV. Considering that diffusion in the Galaxy steepens the spectra during propagation an approximate factor of $E^{-0.3}-E^{-0.5}$, the sum of all sources responsible for this spectrum must have an index of around $E^{-2.2}-E^{-2.4}$ at the source.


\begin{figure}[ht]%
 \begin{center}%
 \includegraphics[width=0.9\linewidth]{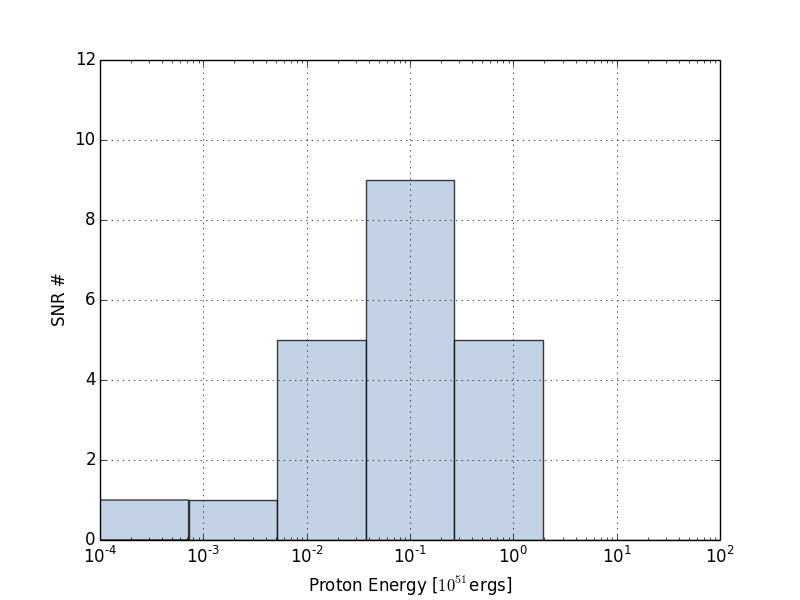}
 \end{center}%
 \caption{Distribution of total energy going into protons, defined in Equ.\ (\ref{etoti:equ}).  \label{etot_p:fig}}
\end{figure}

\begin{figure}[ht]%
 \begin{center}%
 \includegraphics[width=0.9\linewidth]{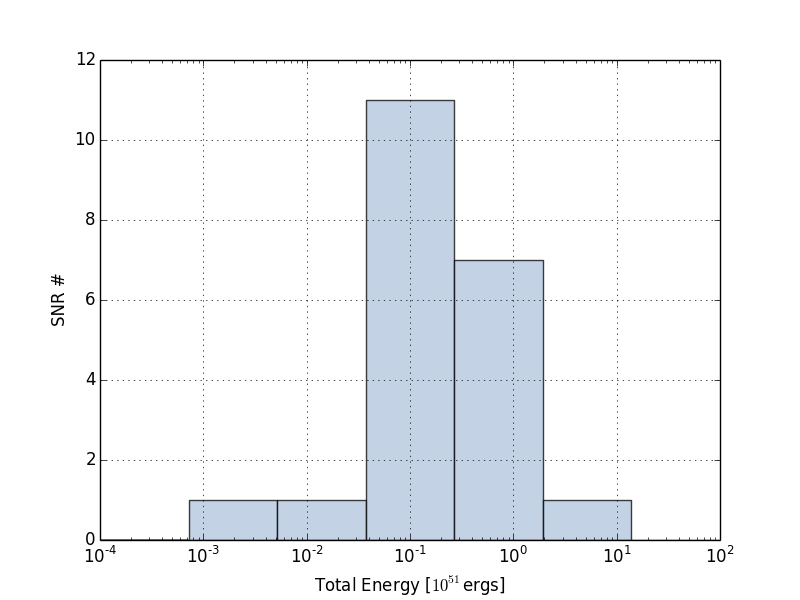}
 \end{center}%
 \caption{Distribution of the total non-thermal energy in an SNR as defined in Equ.\ (\ref{etot:equ})  \label{etot:fig}}
\end{figure}

Figure \ref{etot_p:fig} shows the histogram for the total energy going into protons. The general, the peak is located around $1-2\cdot 10^{50}$~ergs, which is a reasonable value reasonable values, indicating that 10\%-20\% of a standard SN-explosion energy of $10^{51}$~erg goes into cosmic rays. The distribution is somewhat asymmetric with two contributions at $10^{47}-10^{48}$~erg. 
A larger number of sources and improved data for each individual sources will improve the statistics for this histogram: In particular, improved data in the low- and high-energy range will provide a test for the hadronic nature of the sources. At low-energies, the kinematic threshold for proton-proton interactions can be measured as already done for the case of W44 and IC443 \cite{fermi_w44}. At high energies, CTA and HAWC will help to see which sources have spectra persisting up to $100$~TeV - PeV energies. The histogram for the total non-thermal energy of the sources is shown in Fig.\ \ref{etot:fig} and reveals a relatively symmetric distribution with a central value around $3-4\cdot 10^{50}$~erg, which also represent expected values. Another reason could be systematic errors in the assumed distribution of cosmic rays. We actually assume as a first-order approximation, that the energy density we receive from fitting the gamma-ray data is distributed homogeneously over the entire remnant and multiply the energy density by $4/3\cdot \pi\cdot R_{\rm SNR}^{3}$ as filling volume. This is only an approximation, and also relies on a proper estimation of the radius of the remnant $R_{\rm SNR}$, which is difficult. In summary, we believe that there is no problem with the somewhat asymmetric distribution of total proton energy and rather use the graph as a cross-check that the absolute total energy budget per remnant is realistic.

\begin{table}
\centering{
\begin{tabular}{c|ccccccccccc}
\hline\hline
SNR&$B$&$\alpha_e$&$a_e$&$E_{\rm max,e}$&$E_{\rm tot,e}$&$E_{\rm B}$&Refs\\
&[$\mu$G]&&[$10^{39}$ MeV$^{-1}$]&[GeV]&[$10^{47}$~erg]&[$10^{47}$~erg]&SED data\\\hline
3C391 & 128.3 & 1.7 & 215.5 & 1078.4 & 12.1 & 115.8 & \cite{ASCA2001-3C391,Radio1989-Many,FERMI2010-MULT}\\
W41 & 20.4 & 2.2 & 181.7 & 7000.0 & 60.4 & 166.6 & \cite{HESS2006-MULT,Leahy2008-W41,Radio1989-Many,LiChen2012,MAGIC2006-W41,HESS2006-MULT}\\
W33 & 19.8 & 2.0 & 7.2 & 10000.0 & 1.2 & 0.1 & \cite{HESS2006-MULT,MAGIC2006-W33,Radio2005-W33}\\
W30 & 112.1 & 1.9 & 111.8 & 10000.0 & 13.0 & 10840.2 & \cite{Radio1990-W30,FERMI2010-MULT,FERMI2012-W30,HESS2006-MULT}\\
W28 & 11.0 & 1.5 & 21180.5 & 5000.0 & 0.8 & 28570.5 & \cite{Radio1989-Many,HESS2008-W28,FERMI2010-W28}\\
W28C & 44.0 & 2.1 & 1.3 & 5000.0 & 0.2 & 2.3 & \cite{2014ApJ...786..145H,Brogan2006-Many,HESS2008-W28}\\
G359.1-0.5 & 99.4 & 2.2 & 33.3 & 2906.0 & 17.5 & 9040.5 & \cite{FERMI2011-J1745,HESS2006-MULT,HESS2008-J1745,ASCA2000-J1745,2000AJ....119..207L}\\
G349.7+0.2 & 109.9 & 1.9 & 962.0 & 5000.0 & 86.2 & 687.5 & \cite{Radio1989-Many,FERMI2010-MULT}\\
CTB 37B & 58.4 & 2.0 & 822.3 & 5000.0 & 115.5 & 5795.6 & \cite{Radio1991-CTB37AB,HESS2008-CTB37B}\\
CTB 37A & 151.6 & 2.0 & 36.2 & 3494.8 & 4.8 & 9007.8 & \cite{Radio1991-CTB37AB,Suzaku2011-CTB37A,HESS2008-CTB37A,FERMI2010-MULT}\\
RX J1713.7-3946 & 10.4 & 1.9 & 189.6 & 37099.0 & 39.1 & 149.9 & \cite{FERMI2011-RXJ1713,HESS2007-RXJ1713,ATCA2004-RXJ1713,Tanaka2008-RXJ1713}\\
SN 1006 & 60.5 & 2.2 & 3.1 & 9198.2 & 3.2 & 140.4 & \cite{Radio2001-SN1006,2008PASJ...60S.153B,Berezhko2009-SN1006,HESS2010-SN1006}\\
Puppis A & 81.8 & 2.1 & 27.9 & 2210.2 & 9.3 & 1342.9 & \cite{Castelletti2006-PupA,ROSAT1996-PupA,Fermi2012-MULT}\\
Vela Jr & 11.0 & 2.3 & 24.0 & 34150.2 & 86.4 & 80.9 & \cite{Duncan2000,FERMI2011-VelaJr,HESS2007-VelaJr}\\
MSH 11-62 & 18.0 & 1.7 & 57704.6 & 0.0 & 40.4 & 25.6 & \cite{Radio1986-MSH1162,FERMI2012-MSH1162}\\
RCW 86 & 12.7 & 2.3 & 48.7 & 30128.0 & 220.0 & 22.0 & \cite{HESS2009-RCW86,Goumard2012-RCW86,1975AuJPA..37...39C}\\
W44 & 120.1 & 1.7 & 432.3 & 1183.5 & 50.5 & 657.3 & \cite{XMM2006-W44,Castelletti2007-W44,AGILE2011-W44,FERMI2012-W44}\\
G40.5-0.5 & 150.4 & 1.6 & 50.2 & 2500.0 & 2.4 & 1461.5 & \cite{Radio1989-Many,HESS2009-G40.5,MILAGRO2007-G40.5}\\
W49B & 295.4 & 2.5 & 2.2 & 10000.0 & 3.2 & 516.7 & \cite{Radio1989-Many,FERMI2010-W49B,HESS2011-W49B}\\
W51C & 133.3 & 1.4 & 100.9 & 1064.1 & 28.8 & 15611.2 & \cite{FERMI2009-W51C,MAGIC2012-W51C,Moon1994-W51C,ROSAT1995-W51C}\\
IC443 & 70.7 & 1.7 & 195.5 & 50.0 & 2.9 & 698.1 & \cite{FERMI2010-IC443,VERITAS2009-IC443,MAGIC2007-IC443,AGILE2010-IC443,Radio1985-IC443,Milagro2009-Fermi}\\
Cygnus Loop & 60.7 & 2.1 & 4.5 & 2500.0 & 1.5 & 2809.8 & \cite{Radio2004-Cygnus,FERMI2011-Cygnus}\\
Cas A & 100.6 & 2.5 & 25.7 & 14087.5 & 76.4 & 4.2 & \cite{Baars1977-CasA,Araya2010-CasA,FERMI2010-CasA,MAGIC2007-CasA,VERITAS2010-CasA}\\
Tycho & 100.7 & 2.3 & 5.1 & 10923.1 & 15.6 & 33.3 & \cite{Suzaku2012-Tycho,Radio1992-Tycho,FERMI2012-Tycho,VERITAS2011-Tycho}\\
\end{tabular}
\caption{Parameters connected to the electron spectra derived by fitting the SNRs' SED. Column 1 shows the name of the SNR, followed by the magnetic field $B$, and the fit parameters for the electron spectrum, defined Equ.\ (\ref{crs:equ}). The total energies going into electrons and the magnetic field are given in column 6 and 7. References for the SED data are provided in the final column.\label{electrons:tab}}
}
\end{table}

\begin{table}
\centering{
\begin{tabular}{c|cccccccccccccc}
\hline\hline
SNR&$\alpha_{\rm CR}$&$a_{\rm CR}$&$E_{\max,CR}$&$E_{\rm tot,CR}$\\
&&[$10^{39}$ MeV$^{-1}$]&[GeV]&[$10^{47}$~erg]\\\hline
3C391 & 2.6 & 44964.2 & 1000000.0 & 3081.2\\
W41 & 2.4 & 52175.2 & 1000000.0 & 4438.1\\
W33 & 2.1 & 29694.1 & 1000000.0 & 966.0\\
W30 & 2.9 & 19853.4 & 13951.6 & 681.9\\
W28 & 2.8 & 9952.4 & 1000000.0 & 1874.6\\
W28C & 2.5 & 2331.8 & 1000000.0 & 29.3\\
G359.1-0.5 & 0.0 & 0.00 & 0.0 & 0.0\\
G349.7+0.2 & 2.4 & 332128.6 & 1000000.0 & 3155.2\\
CTB 37B & 2.1 & 29721.8 & 1000000.0 & 3745.9\\
CTB 37A & 2.6 & 5835120.6 & 1000000.0 & 1241.3\\
RX J1713.7-3946 & 0.0 & 0.00 & 0.0 & 0.0\\
SN 1006 & 2.3 & 2676.1 & 1000000.0 & 1227.6\\
Puppis A & 2.5 & 4719.8 & 1000000.0 & 231.2\\
Vela Jr & 1.8 & 16348.6 & 43970.1 & 1389.6\\
MSH 11-62 & 1.7 & 2869.8 & 46.0 & 4.2\\
RCW 86 & 0.0 & 0.00 & 0.0 & 0.0\\
W44 & 2.6 & 258.4 & 58.7 & 1.1\\
G40.5-0.5 & 2.0 & 22697.4 & 1000000.0 & 71.2\\
W49B & 2.9 & 76237.4 & 1000000.0 & 1323.3\\
W51C & 2.4 & 118406.8 & 1000000.0 & 7872.5\\
IC443 & 2.7 & 6046.8 & 1000000.0 & 85.2\\
Cygnus Loop & 2.9 & 93.2 & 1000000.0 & 251.9\\
Cas A & 2.3 & 19276.6 & 37315.5 & 2317.8\\
Tycho & 2.3 & 2678.0 & 1000000.0 & 1813.6\\
\end{tabular}
\caption{Parameters connected to the cosmic ray spectra derived by fitting the SNRs' SED. Column 1 shows the name of the SNR, followed by the fit parameters for the hadronic cosmic ray spectrum, defined Equ.\ (\ref{crs:equ}). The maximum energy of the hadronic spectrum is set to $10^{6}$~GeV when no clear cutoff was present in the data. For sources with a cutoff in the data, the maximum energy was kept as a free parameter. The total energy going into hadrons is given in column 5.  Those three sources which have a pure leptonic fit lack hadronic cosmic ray data and are listed as $0.0$ here. \label{hadrons:tab}}
}
\end{table}
\subsection{The role of secondary elecrons and positrons\label{secondaries_result_discussion:sec}}
In the introduction, we specify that we neglect synchrotron radiation from secondary electrons and positrons. The situation is clear for most of the considered remnants, as they have a power in secondary electrons and positrons which is a few orders below the detected synchrotron power.
There are, however, five cases in which the total luminosity in synchrotron radiation is of the same order as the luminosity of secondary electrons and positrons, i.e.\
\begin{eqnarray}
\left(L_{\rm synch,obs},\,L_{\rm esec}\right)_{\rm IC443} &\sim& \left(8\cdot 10^{34}~{\rm erg/s},\,10^{35}~{\rm erg/s}\right)\\
\left(L_{\rm synch,obs},\,L_{\rm esec}\right)_{\rm MSH} &\sim& \left(6\cdot 10^{34}~{\rm erg/s},\,4\cdot 10^{35}~{\rm erg/s}\right)\\
\left(L_{\rm synch,obs},\,L_{\rm esec}\right)_{\rm W28C} &\sim& \left(5\cdot 10^{34}~{\rm erg/s},\,2\cdot 10^{34}~{\rm erg/s}\right)\\
\left(L_{\rm synch,obs},\,L_{\rm esec}\right)_{\rm W33} &\sim& \left( 10^{35}~{\rm erg/s},\,10^{35}~{\rm erg/s}\right)\\
\left(L_{\rm synch,obs},\,L_{\rm esec}\right)_{\rm W49B} &\sim& \left(4\cdot 10^{36}~{\rm erg/s},\,4\cdot 10^{36}~{\rm erg/s}\right)\,.
\end{eqnarray}
These sources could thus be potential candidates to be dominated by secondary electrons, if these particles lose their entire energy to synchrotron radiation. In order to cross-check our assumption, we consider
the average time scale for synchrotron losses at SNRs as given by \cite{parizot2006}
\begin{equation}
\tau_{\rm synch}\approx 2\cdot 10^{3}\,{\rm yr}\cdot \left(\frac{E}{\mbox{TeV}}\right)^{-1}\cdot \left(\frac{B}{100\mu\mbox{G}}\right)^{-2}\,.
\end{equation}
Comparing the synchrotron time scale to other lifetime restricting electrons shows that for the five SNRs above, the age of the remnant is actually the most constraining factor: The remnants considered above are relatively young with ages of below $3000$~years (an exception is W28C, for which the age of the remnant is not known, see Table \ref{basics:tab} for the exact numbers). In order for the electrons to lose all their energy to synchrotron radiation, the synchrotron time scale needs to be shorter than the lifetime of the remnant:
\begin{equation}
\tau_{\rm synch}\approx 2\cdot 10^{3}\,{\rm yr}\cdot \left(\frac{E}{\mbox{TeV}}\right)^{-1}\cdot \left(\frac{B}{100\mu\mbox{G}}\right)^{-2}\ll t_{\rm SNR}\approx2\cdot 10^{3}\,{\rm yr}\,.
\end{equation}
Here, for simplicity, we approximate all remnants with an age of $2000$~years. This means, losses can only be fully effective for
\begin{equation}
 \left(\frac{E}{\mbox{TeV}}\right)\gg \left(\frac{B}{100\mu\mbox{G}}\right)^{-2}\cdot \left( \frac{t_{\rm SNR}}{2\cdot 10^{3}\,{\rm yr}}\right)^{-1}\,.
\end{equation}
The magnetic field strength derived in this approach for the individual remnants is given in Table \ref{electrons:tab}. Inserting the exact numbers from Tables \ref{basics:tab} and \ref{electrons:tab} for age and mangetic field provides us with a necessary condition for fully effective synchrotron loss of\footnote{The value for W28C has been derived by assuming the remnant evolves similarly as W33. We have scaled the lifetime of W33 by the ratio of the remnants' sizes, so that the assumed lifetime of W28C is $t_{\rm SNR,W28C}\approx 2.175\cdot 10^{3}$~yr. }
\begin{equation}
E\gg \left\{\begin{array}{lll}
42.5\,{\rm TeV}&&\mbox{ for W33}\\
5.17\,{\rm TeV}&&\mbox{ for W28C}\\
47.5\,{\rm TeV}&&\mbox{ for MSH}\\
1.33\,{\rm TeV}&&\mbox{ for IC443}\\
5.91\,{\rm TeV}&&\mbox{ for W49B}\\ \end{array}\right.
\end{equation}
For all of these remnants, synchrotron losses are only fully effective above TeV energies\footnote{Even at these energies, the optical depth of the process is smaller than one, so that only part of the luminosity is lost to synchrotron radiation.}. Most of the energy of these electrons is stored at much lower energies, down to sub-GeV energies, so the synchrotron luminosity from secondaries is expected to be much smaller than the luminosity in the secondaries themselves. 
We also calculate the synchrotron radiation from secondary electrons and positrons using the synchrotron loss time scale, and get upper limits for the energy going into synchrotron radiation:
\begin{eqnarray}
\left(L_{\rm synch,obs},\,L_{\rm synch,sec}\right)_{\rm IC443} &\sim& \left(8\cdot 10^{34}~{\rm erg/s},\,3\cdot 10^{33}~{\rm erg/s}\right)\\
\left(L_{\rm synch,obs},\,L_{\rm synch,sec}\right)_{\rm MSH} &\sim& \left(6\cdot 10^{34}~{\rm erg/s},\,2\cdot 10^{31}~{\rm erg/s}\right)\\
\left(L_{\rm synch,obs},\,L_{\rm synch,sec}\right)_{\rm W28C} &\sim& \left(5\cdot 10^{34}~{\rm erg/s},\,2\cdot 10^{31}~{\rm erg/s}\right)\\
\left(L_{\rm synch,obs},\,L_{\rm synch,sec}\right)_{\rm W33} &\sim& \left( 10^{35}~{\rm erg/s},\,10^{33}~{\rm erg/s}\right)\\
\left(L_{\rm synch,obs},\,L_{\rm synch,sec}\right)_{\rm W49B} &\sim& \left(4\cdot 10^{36}~{\rm erg/s},\,3\cdot 10^{32}~{\rm erg/s}\right)\,.
\end{eqnarray}
%
For these five remnants, the ratio between the observed synchrotron radiation and the one expected from secondaries is thus certainly less than $3.8\%$ (the maximum emission for IC443). For all other remnants, the ratio is even much smaller than that.
We conclude that also for these five sources, secondary electrons and positrons can be neglected as the expected output is much smaller than the observed synchrotron luminosities.

\section{Resulting neutrino spectra \label{nu_spectra:sec}}
In this Section, the neutrino spectra received in the hadronic scenario described above are presented for the 21 individual sources (Section \ref{individual:sec}). The derivation of a diffuse neutrino flux from the Galactic Plane is presented in Section \ref{diffuse_nus:sec}.
\subsection{Spectra from individual SNRs\label{individual:sec}}
\begin{figure}[ht]%
 \begin{center}%
 \includegraphics[width=0.9\linewidth]{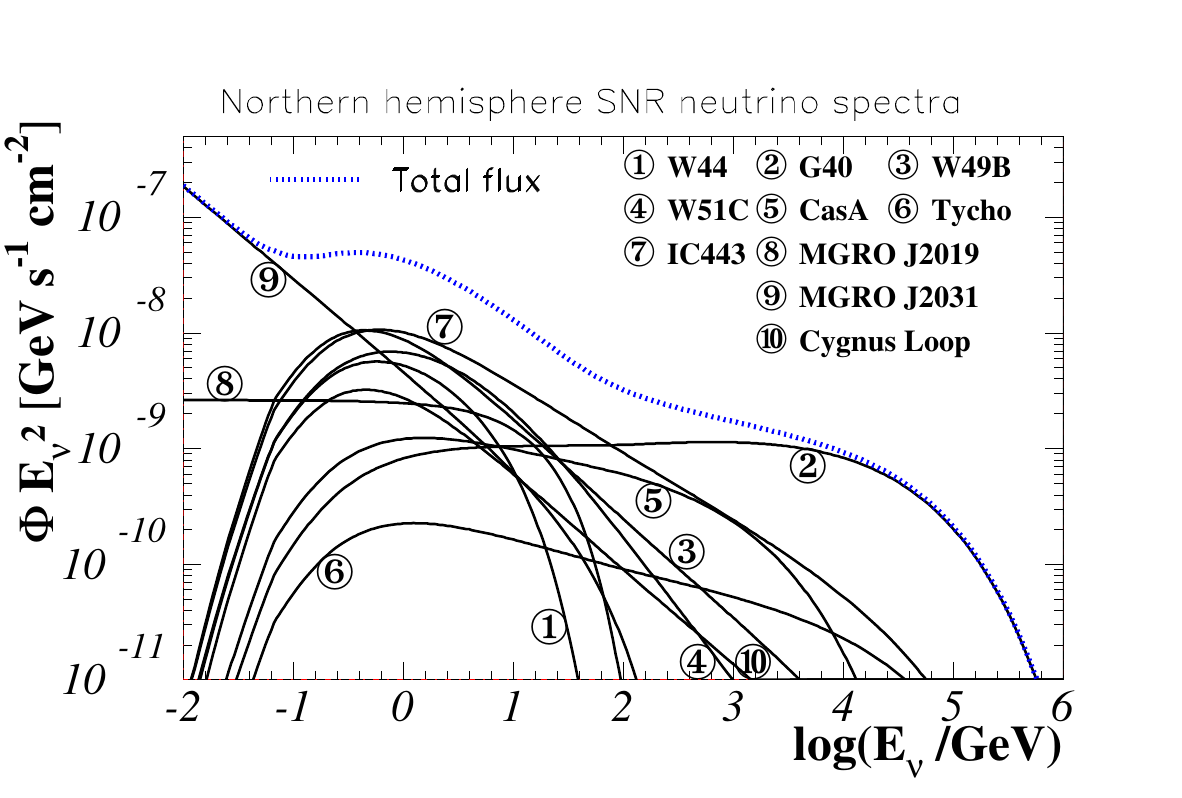}
 \end{center}%
 \caption{Neutrino flux predictions for northern hemisphere Supernova Remnants (black lines). The blue, dotted line shows the sum of all northern spectra. The maximum energy for those spectra that do not show a cutoff in gamma-rays is assumed to be $E_{\rm max, CR}=1$~PeV in this figure. \label{north:fig}}
\end{figure}

The individual neutrino spectra from each SNR are automatically
provided by the hadronic part of the gamma-ray fit, following Equations (\ref{delta_approx:equ}) and (\ref{kelner:equ}).
Figure \ref{north:fig} shows the neutrino flux predictions for northern hemisphere SNRs (black lines). Note that the two Milagro sources do not include the kinematic low-energy cutoff at around 100~MeV and that these spectra are therefore only realistic above 100 MeV. For IceCube, the most relevant contribution is at above the detector threshold at around 100 GeV. Point source searches in the northern hemisphere with IceCube are typically optimized to be sensitive in an energy range of $\sim 1-100$~TeV \cite{icecube_ps2014}. In that range, the strongest sources in the northern hemisphere  are IC443, G40.5-0.5 and CasA. These three sources dominate the total contribution (blue, dotted line) at high energies and the point source search of IceCube is most sensitive to those three sources. Right now, limits are a factor of $\sim 3$ (G40.5-0.5), $\sim 15$ (IC443) and $\sim 30$ (CasA) above the predictions. The best chances for detection within the next few years therefore concerns G40.5-0.5. In \cite{icecube_ps2014}, the authors expect the sensitivity to increase by more than a factor of $2$ within the next four years. Given the fact that IceCube supposedly runs longer than this, IceCube might get close to detection of G40.5+0.5 within its lifetime.

\begin{figure}[ht]%
 \begin{center}%
 \includegraphics[width=0.9\linewidth]{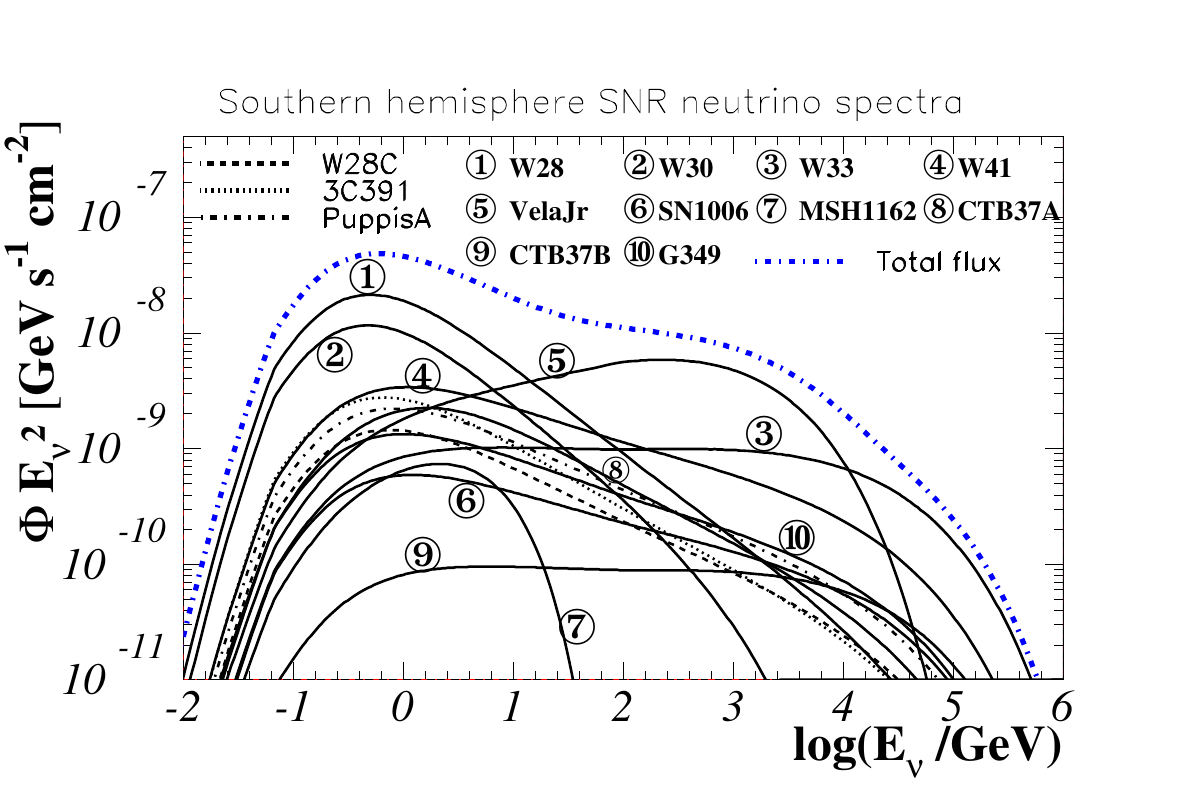}
 \end{center}%
 \caption{Neutrino flux predictions for southern hemisphere Supernova Remnants (black lines). The blue, dot-dashed line shows the sum of all southern spectra. The maximum energy for those spectra that do not show a cutoff in gamma-rays is assumed to be $E_{\rm max, CR}=1$~PeV in this figure.  \label{south:fig}}
\end{figure}

Figure \ref{south:fig} displays the neutrino flux predictions for southern hemisphere SNRs (black lines). The strongest sources above $100$~GeV are Vela Junior, W33 and W41. These three sources make up the dominant part of the flux above $100$~GeV. This southern hemisphere flux from photon-resolved SNRs is about a factor of $3-4$ larger than the one in the northern hemisphere (see blue, dot-dashed line), which is expected due to the enhanced star-forming activity in the Galactic center region. An exception in the northern hemisphere is the Cygnus region, which also shows strong star-forming activity. The most prominent sources in this region are included in this analysis, there might still be additional contribution from so far unidentified sources. IceCube point sources searches in the southern hemisphere are, however, only sensitive to a signal above $100$~TeV and as the fluxes already start to decrease due to the high-energy cutoff of cosmic rays, southern hemisphere SNRs are not really accessible for the point source search in IceCube. They might still contribute in other channels that are not as sensitive to direction, but have a lower energy threshold even in the southern hemisphere (see discussion of the diffuse flux prediction). Antares, located in the northern hemisphere, is sensitive to these sources at a level of $E^2dN/dE \sim 6\cdot 10^{-8}$~GeV/(s cm$^{2}$) below 100 TeV \cite{antares_ps2014}, which is a factor of $>6$ above the predicted flux of Vela Junior and a factor of $< 100$ above W33 and W41. The given sensitivity is for an $E^{-2}-$type flux and is expected to be worse for other type of spectra, so in the case of the sources here, the real factor is expected to be even larger. Thus, at this point, these sources are not expected to provide a significant point source signal. In the future, KM3NeT is expected to reach the sensitivity level for at least Vela Junior \cite{km3net}. 

\subsection{Derivation of the diffuse neutrino flux from SNRs\label{diffuse_nus:sec}}

As described above, out of 24 well-studied SNRs, $21$ can be modeled hadronically. For completeness, we add two sources from the Cygnus region as described in \cite{gonzalez_garcia2014} to our sample, so that a total of $n_{\max}=23$ sources is included, 10 being in the northern and 13 being in the southern hemisphere. In order to receive the quasi-diffuse neutrino flux from these resolved SNRs, we sum over all individual point sources fluxes $\Phi_i$ and divide by the area of the sky which is covered by the Galactic plane, i.e.\ about 10\% of the sky,  $\pi/3$:
\begin{equation}
\diffnures = \frac{\sum_{i=1}^{n_{\max}=23}\Phi_{i}}{\pi/3}
\end{equation}
Note that when calculating the total number of neutrinos in a neutrino telescope, one has to account for the field of view of the detector and the part of the Galactic plane visible in the FoV, so generally only a fraction of the visible sky.

\begin{figure}[ht]%
 \begin{center}%
 \includegraphics[width=0.9\linewidth]{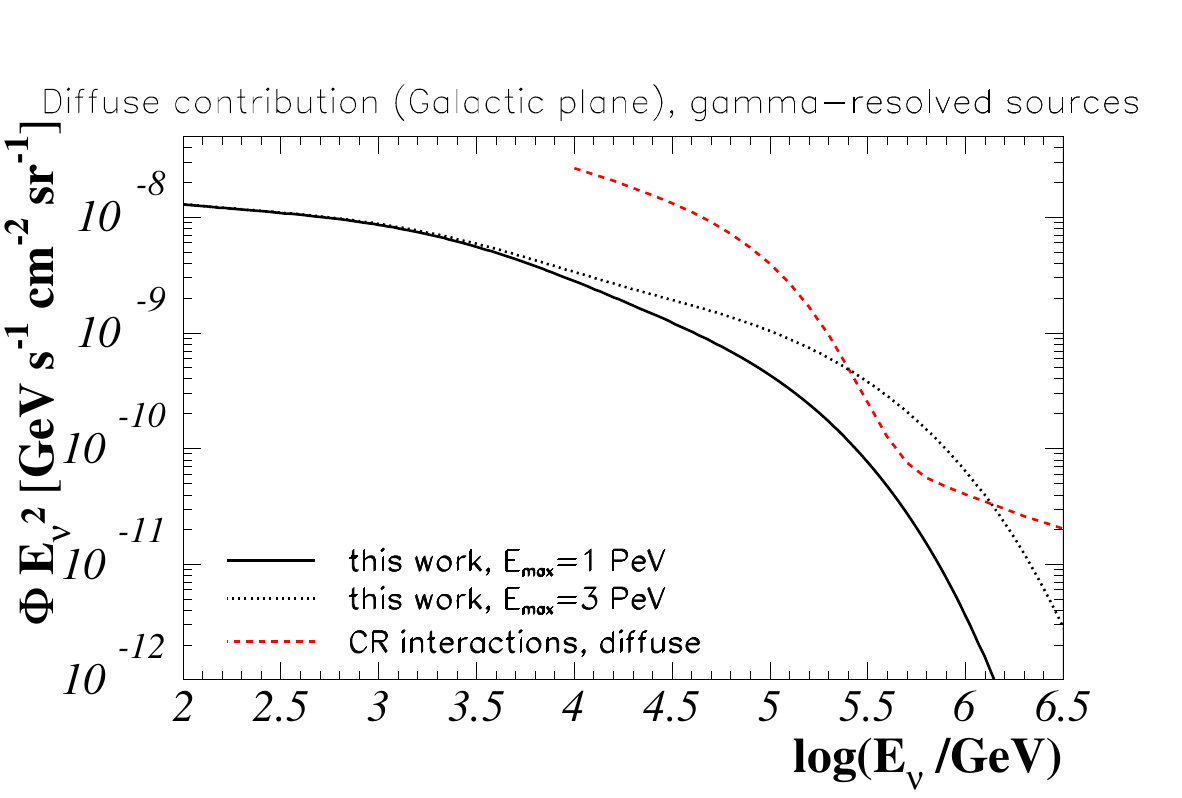}
 \end{center}%
 \caption{Neutrino flux from resolved gamma-ray SNRs - the solid line shows the prediction for a high-energy cutoff at $E_{\rm max, CR}=1$~PeV for those sources that do not show a cutoff in the gamma-ray spectra, while the dashed line represents $E_{\rm max, CR}=3$~PeV in this figure. We compare our results to the prediction of the diffuse neutrino flux from cosmic ray interactions in the interstellar medium \cite{winter_galactic2014} (red, dashed line). This recent model predicts a rather low flux compared with previous models (see \cite{learned_mannheim} for a review). \label{diffuse_emax:fig}}
\end{figure}

The diffuse flux including the contribution from all 23 resolved SNRs is shown in Fig.\ \ref{diffuse_emax:fig}. The predicted flux is shown for two different maximum energies for those sources that do not show a cutoff at gamma-ray energies: the solid line represents $E_{\rm max, CR}=1$~PeV, while the dashed line is the prediction for $E_{\rm max, CR}=3$~PeV. The cosmic ray knee is observed at around 1 PeV, but systematic uncertainties in the energy scale might allow for a slight shift. Also, there could be individual remnants that accelerated to higher energies.

From Green's catalog, we know that there exist close to 300 SNRs in the Galaxy, which also should contribute to the total quasi-diffuse neutrino flux.
In order to estimate this total, diffuse neutrino flux, we use the radio data, which for the gamma-ray detected SNRs correlate with the high-energy signature as shown above.
In a first step, we use the radio data at $1$~GHz to determine the flux for each individual, gamma-ray detected remnant $F_{\rm radio}^{i}$ and sum over all remnants, $F_{\rm radio}^{\rm res}=\sum_{i=1}^{n_{\max}=23}F_{\rm radio}^{i}$. The two Milagro sources, which we could not fit individually due to the lack of multiwavelength data were taken into account as follows: For MGRO2031+41, the radio source G106.6+2.9 appears as the most reasonable counterpart. The measured radio flux above $4.84$~GHz indicates a spectral behavior of $\nu^{-0.77}$ \cite{kothes2006} and the radio flux at $1$~GHz can thus be derived to be around $F_{\rm radio}(1$~GHz$)\approx25$~Jy.  No radio counterpart could be identified for MGRO2019+37 and is therefore considered to be negligible.  The total radio flux at $1$~GHz for the gamma-ray resolved sources considered in this paper is $F_{\rm radio}^{\rm res}\approx 3220$~Jy.

\begin{figure}[ht]%
 \begin{center}%
 \includegraphics[width=0.9\linewidth]{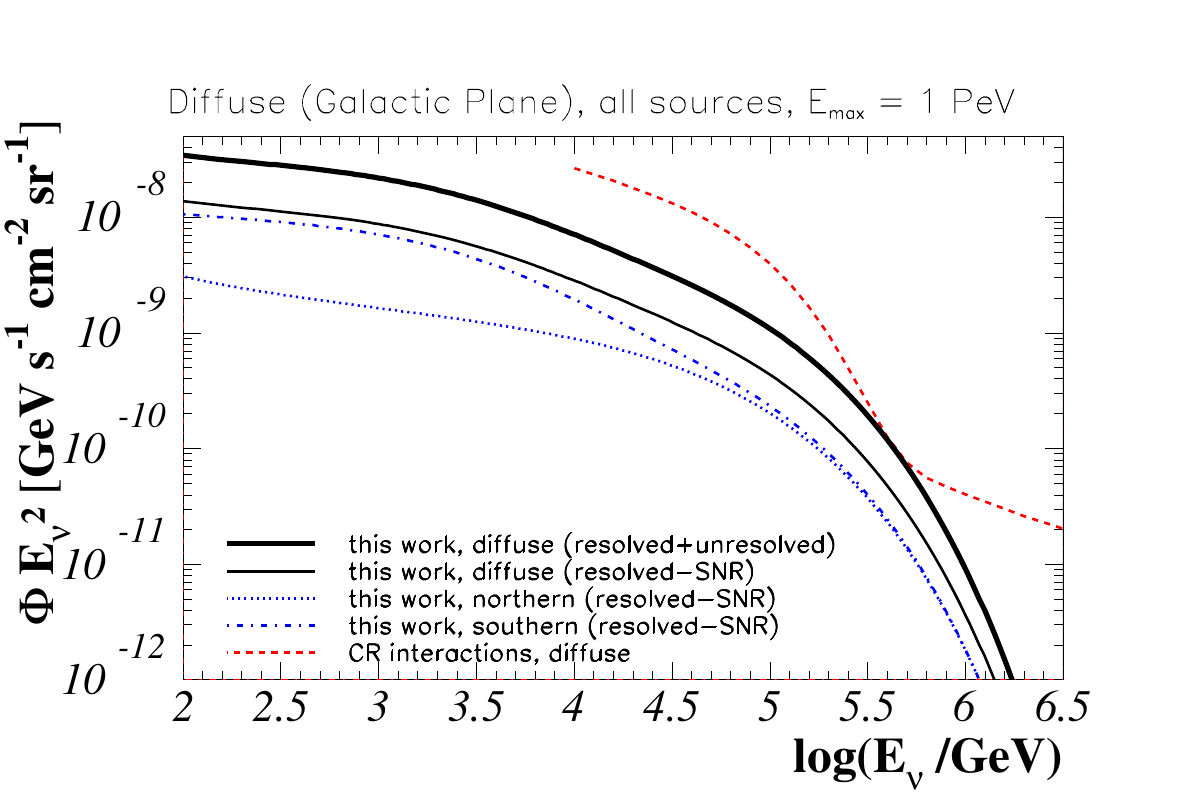}
 \end{center}%
 \caption{Diffuse neutrino flux from
   SNRs in the Milky Way in
   comparison to the neutrino
   emission in the Galaxy from
   cosmic ray interactions in the
   interstellar medium
   \cite{winter_galactic2014}. The
   maximum energy for those sources
   not showing a high-energy cutoff
   in gamma-rays is chosen to be
   $1$~PeV in this graph.\label{diffuse:fig}}
\end{figure}

\begin{figure}[ht]%
 \begin{center}%
 \includegraphics[width=0.9\linewidth]{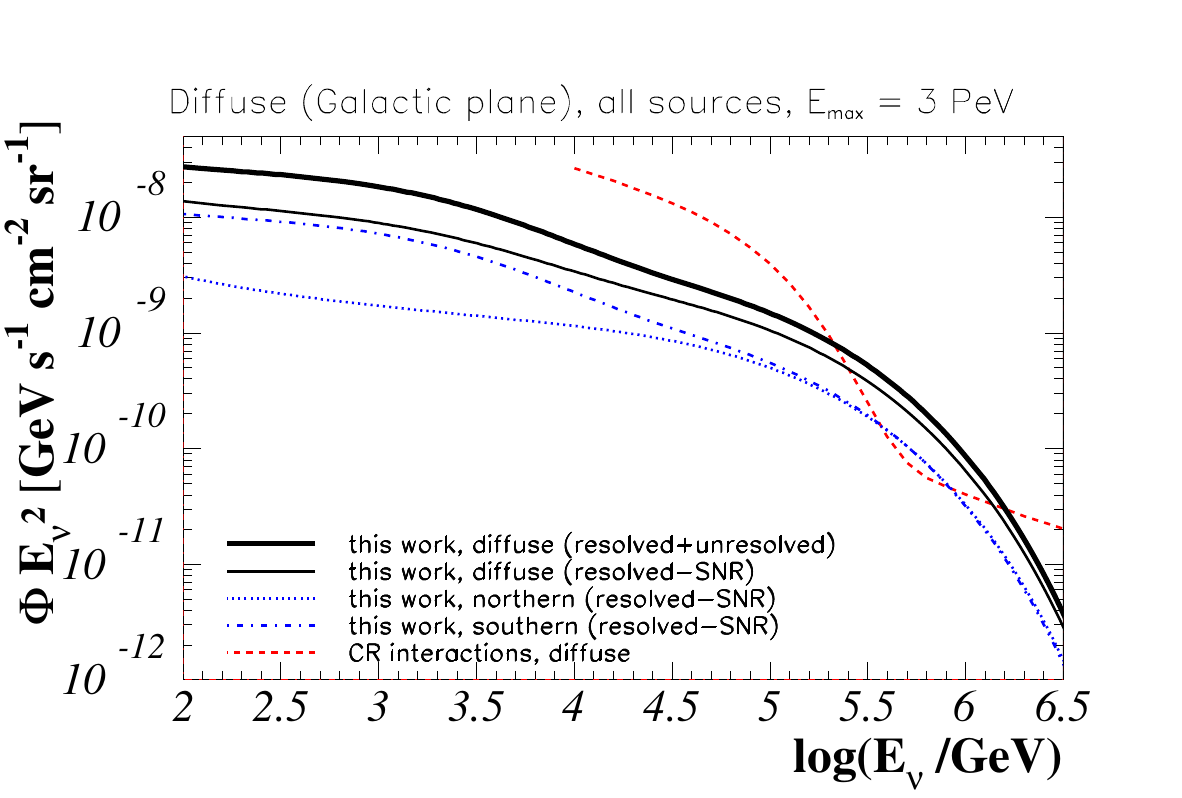}
 \end{center}%
 \caption{Diffuse neutrino flux from
   SNRs in the Milky Way in
   comparison to the neutrino
   emission in the Galaxy from
   cosmic ray interactions in the
   interstellar medium
   \cite{winter_galactic2014}. The
   maximum energy for those sources
   not showing a high-energy cutoff
   in gamma-rays is chosen to be
   $3$~PeV in this graph.\label{diffuse_3pev:fig}}
\end{figure}

Now, we can estimate the total flux of gamma-ray unresolved sources by assuming that $F_{\gamma}\propto F_{\rm radio}$ (see Section \ref{nu_gamma:sec}). For the radio flux of gamma-ray unresolved sources, we use Green's catalog. Here, 274 SNRs are listed. In order to  derive the total flux from gamma-ray unresolved sources, we remove those 21 SNRs that are already included in our sample\footnote{W28C and Cygnus Loop are not part of Green's catalog, so they do not need to be removed. The W51 complex is included as one measurement in Green's catalog. In our sample W51C is listed separately and we subtract our measured flux of $35$~Jy from the total flux given for W51, $160$~Jy, resulting in a remaining flux for W51 of $125$~Jy.}. We further remove 10 SNRs from Green's catalog, for which the emission fills the SNR, so that it is likely that the radio signal comes from the Pulsar Wind Nebula rather than the SNR shell. The Crab nebula, for example, is among those sources and is removed this way as an unlikely hadronic source. An additional 9 sources are removed due to the lack of radio data (they are identified at other wavelengths). Also here, we assume that the contribution can be neglected. Summing up the radio fluxes at $1$~GHz from the remaining 234 sources yields a contribution of $F_{\rm radio}^{\rm unres}\approx 4860$~Jy. The ratio between unresolved and resolved sources is therefore 
\begin{equation}
\frac{F_{\rm radio}^{\rm unres}}{F_{\rm radio}^{\rm res}}\approx 1.5\,.
\label{radio_ratio:equ}
\end{equation}
In order to calculate the resulting neutrino flux $\diffnu$, we now assume that the total, diffuse flux from SNRs is proportional to the sum of the resolved and unresolved sources:
\begin{equation}
\diffnu=\diffnures+\diffnuunres=\left(1+\frac{\diffnuunres}{\diffnures}\right)\cdot \diffnures\,.
\end{equation}
With a proportionality between radio and hadronic signal, we find
\begin{equation}
\frac{\diffnuunres}{\diffnures}=\frac{F_{\gamma}^{\rm unres}}{F_{\gamma}^{\rm res}}=\frac{F_{\rm radio}^{\rm unres}}{F_{\rm radio}^{\rm res}}
\end{equation}
and thus
\begin{equation}
\diffnu=\left(1+\frac{F_{\rm radio}^{\rm unres}}{F_{\rm radio}^{\rm res}} \right)\cdot \diffnures\,.
\end{equation}
Using equation \ref{radio_ratio:equ}, the total diffuse neutrino flux from SNRs can be derived from the total flux of resolved sources:
\begin{equation}
F_{\nu}^{\rm diffuse}\approx 2.5\cdot F_{\nu}^{\rm res}\,.
\end{equation}
This implies that a large fraction (150\%) of the total, diffuse neutrino flux from cosmic ray interactions near SNRs is still unresolved.
Figure \ref{diffuse:fig} presents the prediction for the total diffuse neutrino flux from resolved and unresolved sources. Here, we assume that the spectral behavior of the resolved sources is representative for the unresolved sources as well and we apply a factor of 2.5 to account for unresolved sources as described above. The blue lines represent the contribution from the northern (dotted) and southern (dot-dashed), resolved sources. The thin, solid, black line is the contribution from all resolved SNRs and the thick, solid, black line is the total quasi-diffuse flux from all SNRs in the Galaxy. The total, quasi-diffuse contribution is compared to the intensity of the neutrino flux from cosmic ray interactions in the interstellar medium, as recently calculated in \cite{winter_galactic2014}. Note that this recent calculation of cosmic ray interactions with the ISM is lower than what was predicted previously, see e.g.\ \cite{learned_mannheim} for a review.  At $10-100$~TeV, the contribution from cosmic ray interactions close to SNRs is a factor of $3-4$ lower than for interactions in the diffuse interstellar medium. Above $100$~TeV, both contributions are of comparable intensity and should contribute equally to a total diffuse, Galactic emission. As we do not take into account any possible emission scenario above the knee, the region above $\sim 1$~PeV is not covered by our predictions.
As the gamma-ray spectra in four of the cases do not show signs for a cutoff in the data, we apply two different cases for the maximum energy of primary particles: While Fig.\ \ref{diffuse:fig} displays the spectra for $E_{\rm CR, max}=1$~PeV,  Fig.\ \ref{diffuse_3pev:fig} represents the more optimistic case of $E_{\rm CR, max}=3$~PeV. In this case, the contribution from interactions in the local vicinity of SNRs would even be stronger than the one from interactions in the interstellar medium.

\begin{figure}[ht]%
 \begin{center}%
 \includegraphics[width=0.9\linewidth]{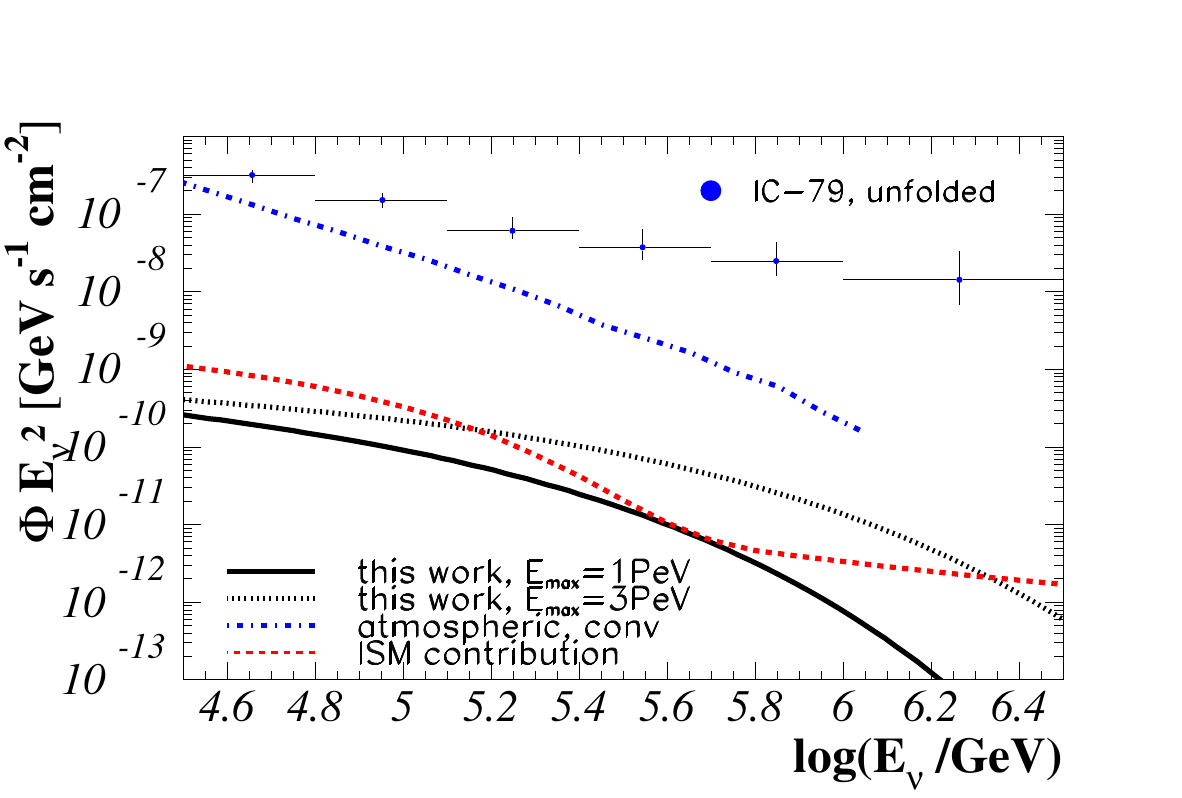}
 \end{center}%
 \caption{Quasi-diffuse neutrino flux from SNRs in the Milky Way in comparison to the neutrino emission in the Galaxy from cosmic ray interactions in the interstellar medium \cite{winter_galactic2014} and the IceCube energy spectrum, revealing an astrophysical component (blue data points) \cite{martin_neutrino2014,florian_ecrs2014} over the conventional background of atmospheric neutrinos (dot-dashed, blue line) \cite{fedynitch2012}.\label{diffuse_data:fig}}
\end{figure}

Figure \ref{diffuse_data:fig} shows the comparison of the quasi-diffuse fluxes with the prediction of the conventional atmospheric neutrino flux (contribution from $\pi,\,K$ particles in air showers) \cite{fedynitch2012} and to the unfolded energy spectrum as measured with IceCube in the 79-strings configuration \cite{martin_neutrino2014,florian_ecrs2014}. For this purpose, as data and atmospheric predictions are averaged over $4\pi$, we smear out the diffuse neutrino flux from the Galactic plane over $4\pi$ as well, leading to a reduction of the flux by a factor $(\pi/3)/(4\pi)=1/12$. In the IceCube-79 data, a deviation from the atmospheric, conventional flux was observed for the first time  and can be taken as a measure for the astrophysical excess, consistent with what was reported previously \cite{icecube2013,icecube2014}. We show predictions for $E_{\rm CR,max}=1$~PeV (solid line) and $E_{\rm CR,max}=3$~PeV (dotted line) as well as the neutrino flux from cosmic ray interactions in the ISM (red, dashed line) \cite{winter_galactic2014}. First of all, it is obvious that the flux is much lower than the conventional atmospheric flux, i.e.\ about one order of magnitude at $1$~PeV assuming the optimistic scenario with $E_{\rm CR, max}=3$~PeV. For a lower energy cutoff at $E_{\rm CR, max}=1$~PeV, the atmospheric flux is about a factor 100 higher at one PeV. At around 30~TeV, where Galactic sources in principle could be contributing to the 37 events reported in \cite{icecube2014}, the predicted neutrino flux is at least a factor of 20 below the signal. We therefore conclude that the diffuse flux derived from well-resolved gamma-ray sources cannot contribute significantly to the neutrino signal detected with IceCube. A dedicated analysis searching for neutrino-induced muon tracks from the direction of the Galactic plane was performed with AMANDA \cite{amanda_gp2005}. The limit derived from four years AMANDA-II data is $E^2\cdot dN/dE = 4.8\cdot 10^{-4}\cdot (E/$GeV$)^{-0.7}$~GeV/(s cm$^2$ sr). At 10~TeV, this limit for an $E^{-2.7}$ flux is therefore around $E^2\cdot dN/dE =10^{-6}$~GeV/(s cm$^2$ sr). With several years of IceCube and a much flatter spectrum which we predict, it is expected that this limit can be improved by several orders of magnitude. Only a dedicated analysis can tell if the sensitivity would be sufficient to observe such a flux. With ANTARES in the mediterranian as well as its successur KM3NeT, a high sensitivity to the southern part of the spectrum can be acheived.

\section{Summary and Conclusions\label{conclusions:sec}}
In this paper, we fit multiwavelength data of 24 supernova remnants which have been identified at $>$~GeV energies. We fit the data including both bremsstrahlung, Inverse Compton scattering and hadronic emission and test a variety of magnetic fields. Finally, for each SNR, we chose a magnetic field which is low enough to keep to total non-thermal energy budget below $10^{51}$~erg within the errors of the calculation
and which is still hadronically dominated at high energies. This
approach works for 21 of the 24 sources. From the hadronic part of the
gamma-ray spectrum, we then derive the corresponding neutrino flux. In
the northern hemisphere, the sources G40.5-0.5, IC443 and CasA are the
strongest ones in the TeV range. In the future, IceCube might be able
to detect G40.5-0.5 as a point source, while the other two sources are
too dim individually. For the case of southern hemisphere sources,
Vela Junior, W33 and W41 are the strongest sources and significantly below the Antares detection threshold. KM3NeT could be able to detect the strongest source Cygnus Loop.

We further derive the diffuse neutrino flux from supernova remnants in the Galactic plane, using the fact that the gamma-ray emission correlates with the radio flux at 1~GHz. Using a well-defined sub-sample of 234 sources from Green's catalog, we show that the total diffuse neutrino flux in the Galaxy lies a factor of $\sim 2.5$ above the flux of so far resolved sources. We show here that the diffuse flux from interactions close to SNRs are of a comparable level as the one from interactions in the interstellar medium and could even supersede it with a high enough cutoff. We find, comparable to what is found for the diffuse neutrino flux from interactions with the ISM \cite{winter_galactic2014}, that the contribution to the signal detected by IceCube is small, as it is at least a factor 20 below the measured flux. Possibly, a dedicated analysis of the Galactic plane could reach a sensitivity comparable to the expected flux. Including unidentified TeV sources will increase the intensity of the flux, but this also increases the uncertainty if the signal is really of hadronic nature as well.

\section*{Acknowledgments}

We thank the anonymous referee for very helpful comments. We also thank Peter Biermann, Dominik
Bomans, Ralf-J\"urgen Dettmar, Jay Gallagher, Francis Halzen,
Bon-Chul Koo, Athina Meli, Wolfgang Rhode, Isaac Saba, Reinhard Schlickeiser, Florian Schuppan, Walter Winter, Marek Wezgowiec
 and Tova Yoast-Hull for very useful discussions. A special thanks to Miguel Araya-Arguedas, Evgeny Berezhko, and Takaaki Tanaka as they provided the X-Ray spectra for SN 1006, Cassiopeia A, and RX J1713.7--3946. We
acknowledge the support from the DFG
research group ``Instabilities, turbulence and transport in
cosmic magnetic fields'' (FOR1048). Further support comes from the MERCUR Project Pr-2012-0008 and from the Research Department of Plasmas
with Complex Interactions (Bochum).
\clearpage
\appendix
\section{Non-thermal energy budget}
\begin{figure}[ht!]%
 \begin{center}%
 \includegraphics[width=\linewidth]{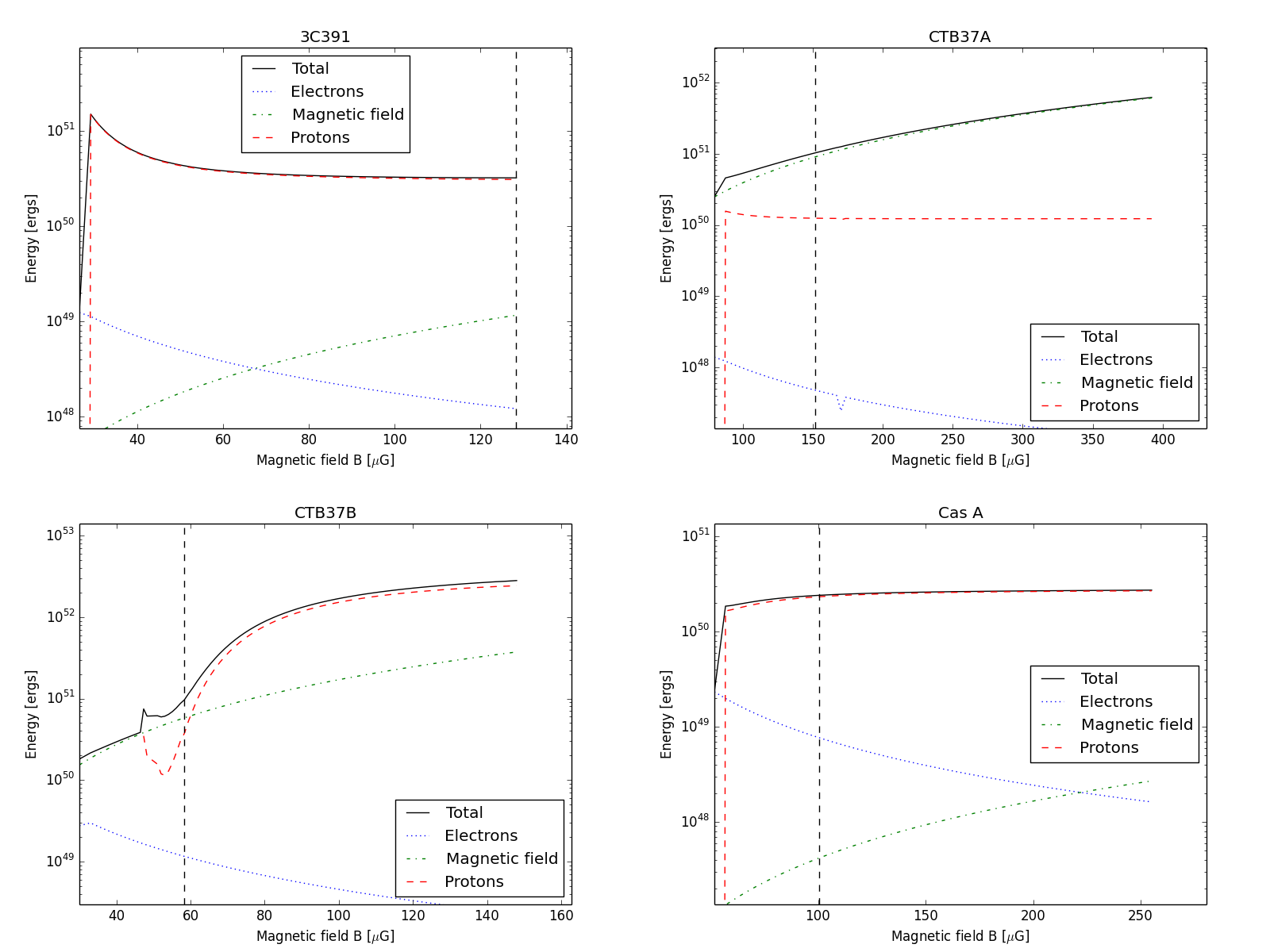}
 \end{center}%
 \caption{Non-thermal energy budget
   for the sources 3C391, CTB37A, CTB37B and Cas A. The
   red, dashed line represents the
   budget of cosmic ray protons,
   the blue, dotted line shows the
   energy of the electrons and the
   green, dot-dashed line displays
   the magnetic field budget at a
   given magnetic field. The black
   line is the sum of all three
   contribution, i.e.\ the total
   non-thermal energy budget. The
   vertical, black, dashed line
   shows the magnetic field value
   chosen in order to model a
   hadronic scenario in which the
   high-energy bump in the photon
   SED is mainly described by
   $\pi^{0}-$ decays. The SEDs
   corresponding to this indicated
   value are shown in Fig.\ \ref{sed0:fig}.    \label{etot0:fig}}
\end{figure}%
\begin{figure}[ht]%
 \begin{center}%
 \includegraphics[width=\linewidth]{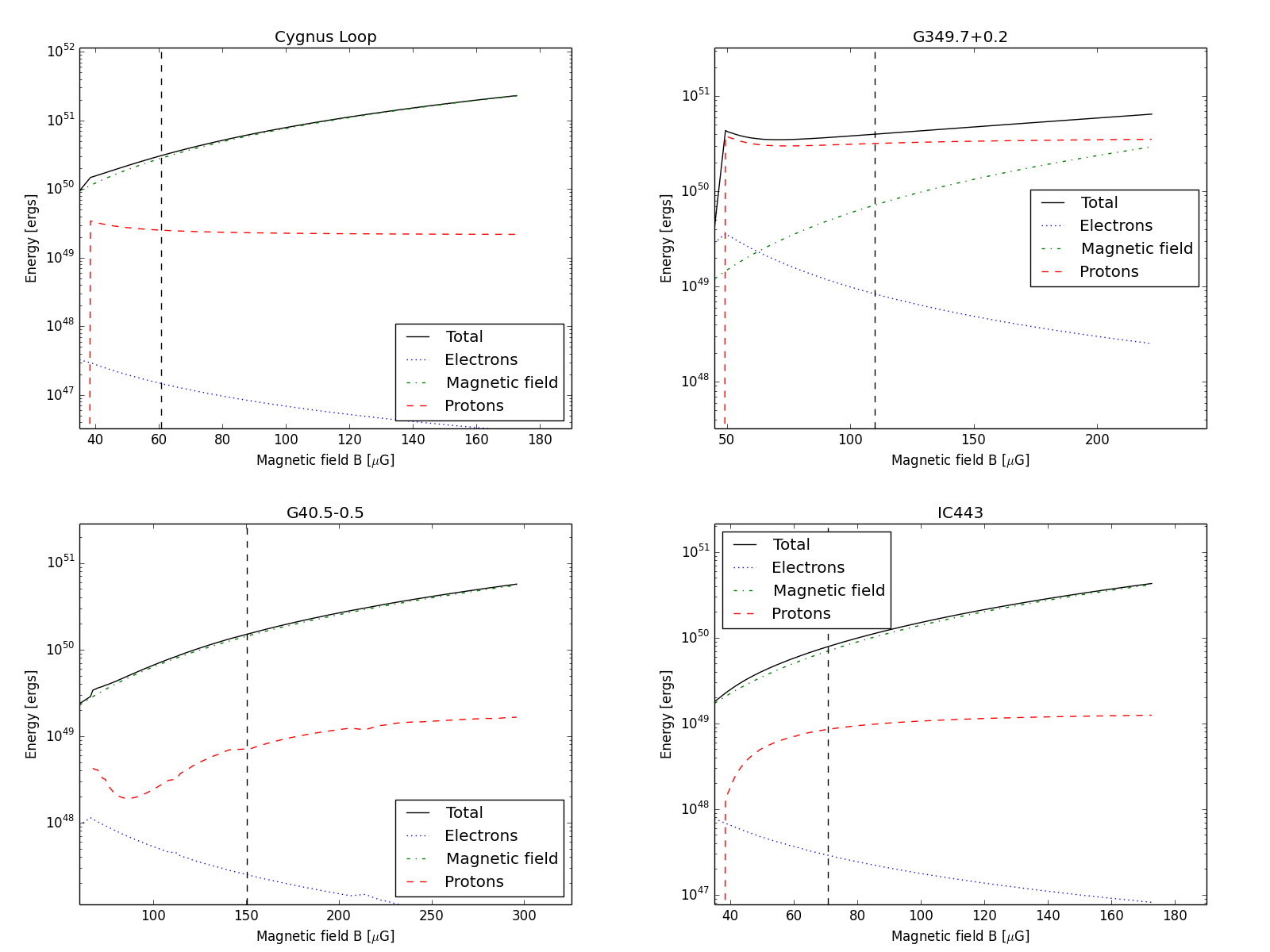}
 \end{center}%
 \caption{Non-thermal energy budget
   for the sources Cygnus Loop, G349+0.2, G40.5-0.5 and IC443. Labeling as in \ref{etot0:fig}. The SEDs
   corresponding to the vertical
   line, chosen as the hadronic scenario, are shown in Fig.\ \ref{sed1:fig}.   \label{etot1:fig}}
\end{figure}%
\begin{figure}[ht]%
 \begin{center}%
 \includegraphics[width=\linewidth]{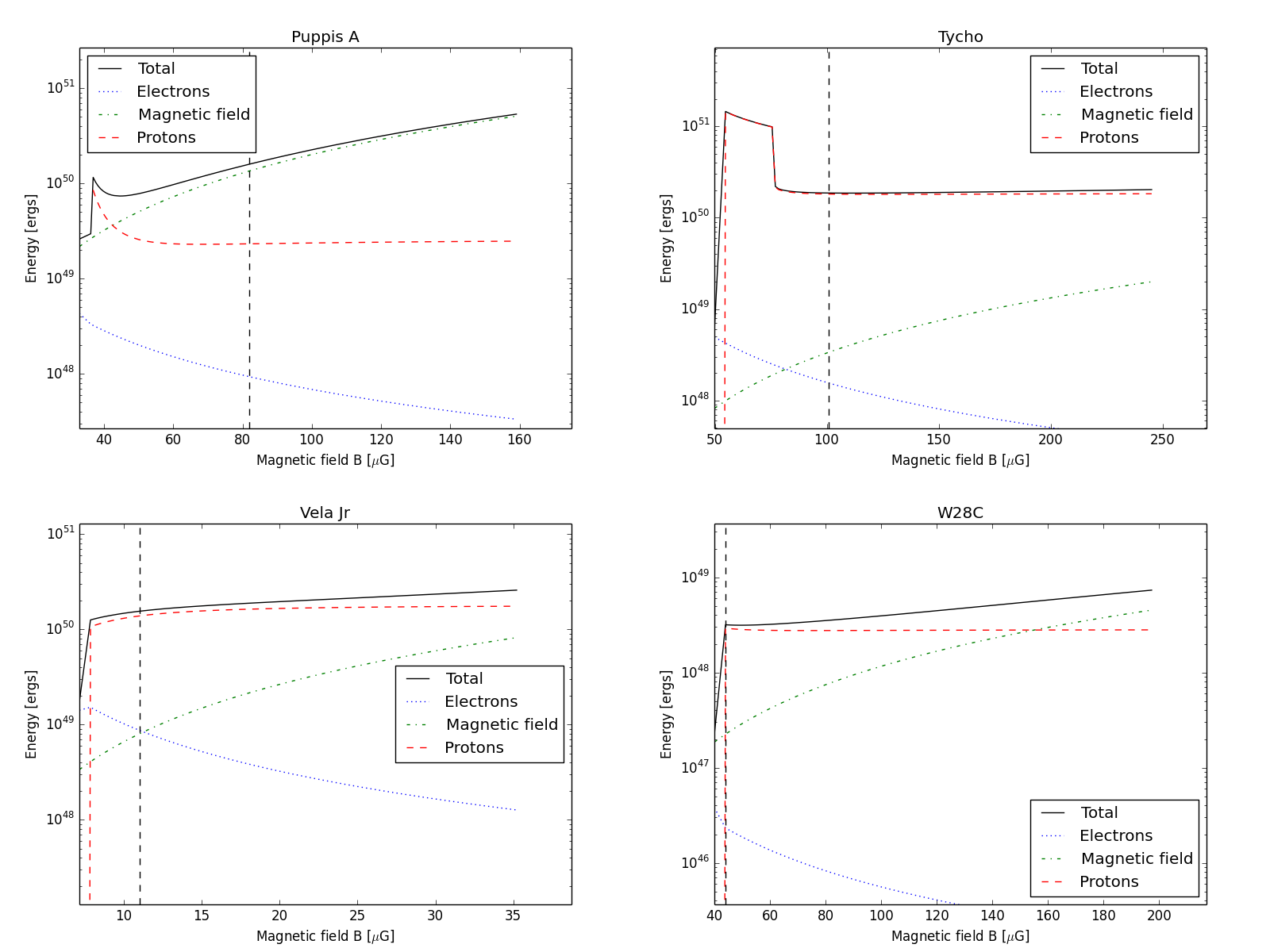}
 \end{center}%
 \caption{Non-thermal energy budget
   for the sources Puppis A, Tycho, Vela Junior and W28C. Labeling as in \ref{etot0:fig}. The SEDs
   corresponding to the vertical
   line, chosen as the hadronic
   scenario, are shown in
   Fig.\ \ref{sed2:fig}.
   \label{etot2:fig}}
\end{figure}%
\begin{figure}[ht]%
 \begin{center}%
 \includegraphics[width=\linewidth]{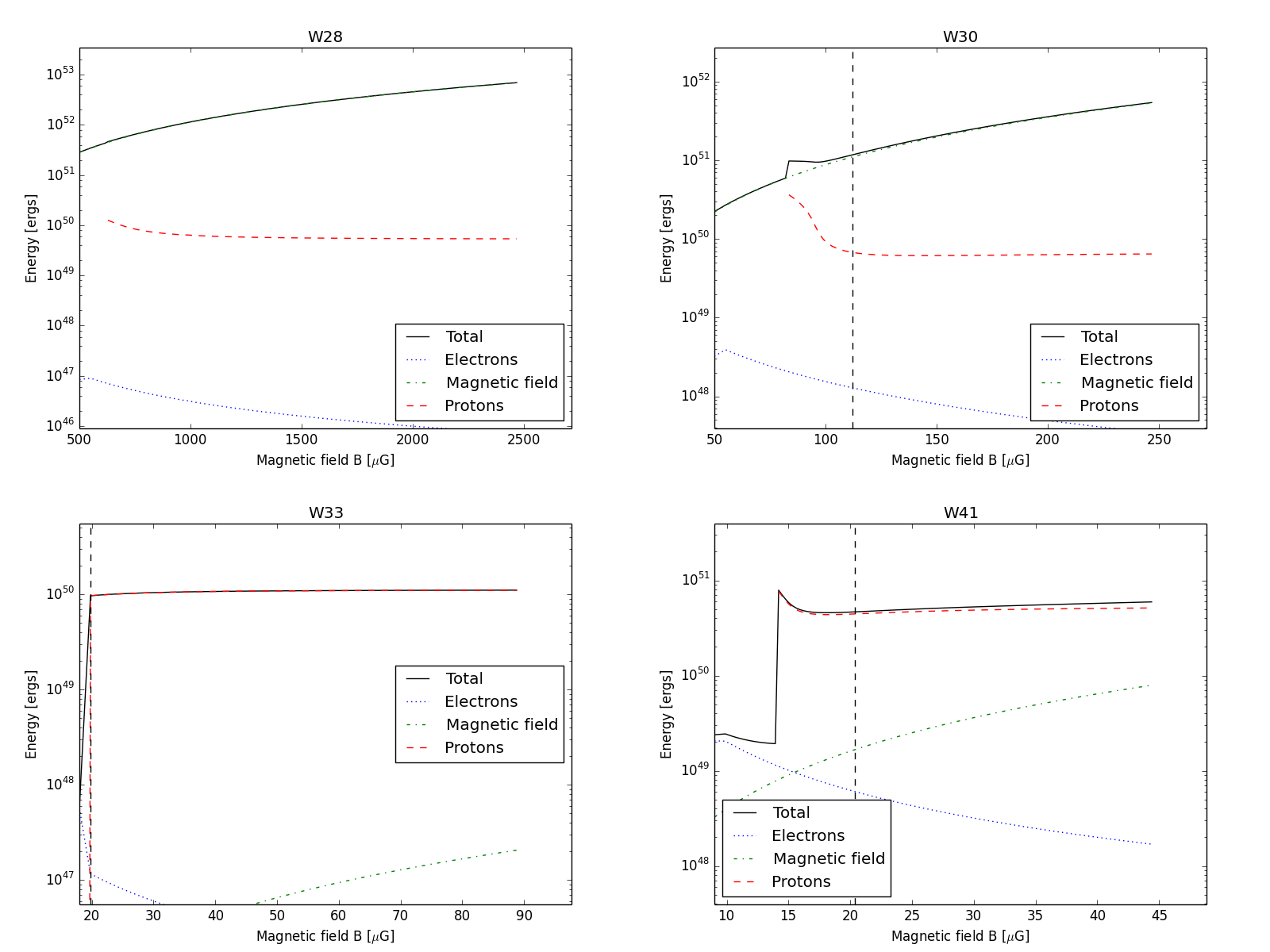}
 \end{center}%
 \caption{Non-thermal energy budget
   for the sources W28, W30, W33 and W41. Labeling as in \ref{etot0:fig}. The SEDs
   corresponding to the vertical
   line, chosen as the hadronic
   scenario, are shown in
   Fig.\ \ref{sed3:fig}.   \label{etot3:fig}}
\end{figure}%
\begin{figure}[ht]%
 \begin{center}%
 \includegraphics[width=\linewidth]{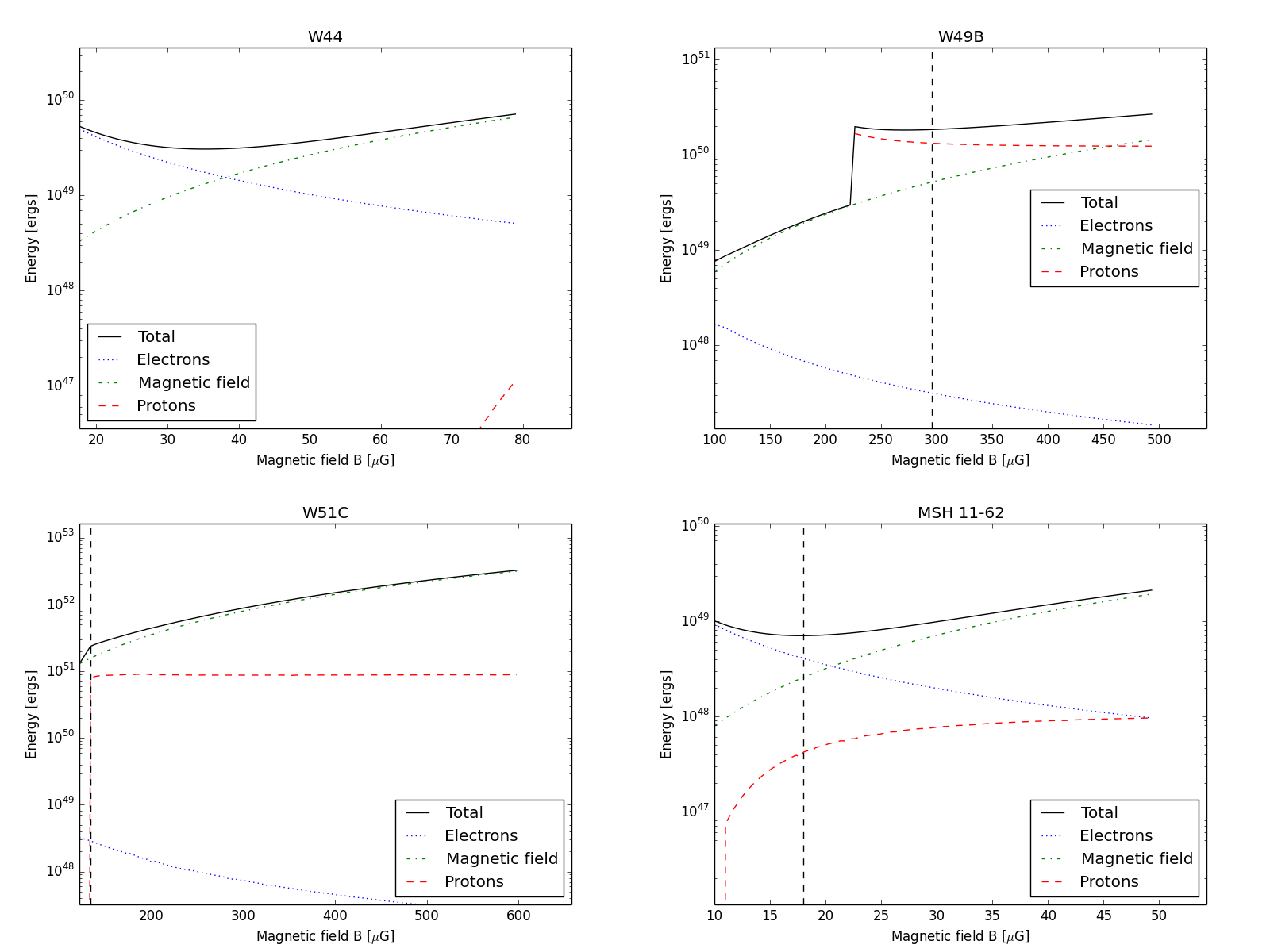}
 \end{center}%
 \caption{Non-thermal energy budget
   for the sources W44, W49B, W51C and MSH 11-62. Labeling as in \ref{etot0:fig}. The SEDs
   corresponding to the vertical
   line, chosen as the hadronic
   scenario, are shown in
   Fig.\ \ref{sed4:fig}. An exception is W44, for which the systematically simulated magnetic field range did not suffice and the hadronic contribution is not saturated yet. Here, we chose $B=120\,\mu$Gauss as the magnetic field, not shown in this plot.  \label{etot4:fig}}
\end{figure}%
\begin{figure}[ht]%
 \begin{center}%
 \includegraphics[width=\linewidth]{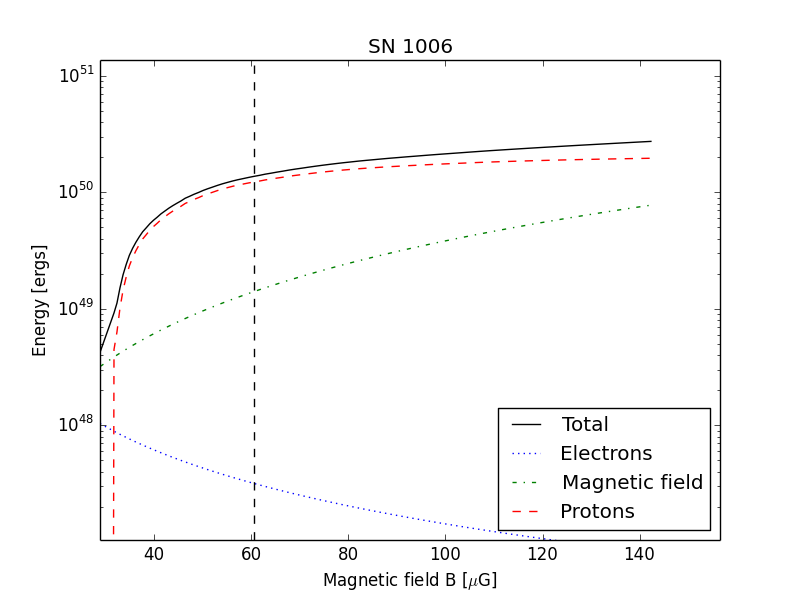}
 \end{center}%
 \caption{Non-thermal energy budget
   for the source SN1006. Labeling as in \ref{etot0:fig}. The SED  corresponding to the vertical
   line, chosen as the hadronic
   scenario, is shown in
   Fig.\ \ref{sed_sn1006:fig}.   \label{etot_sn1006:fig}}
\end{figure}%
\clearpage
\section{SEDs for hadronically
  dominated case}
\begin{figure}[ht!]%
 \begin{center}%
 \includegraphics[width=\linewidth]{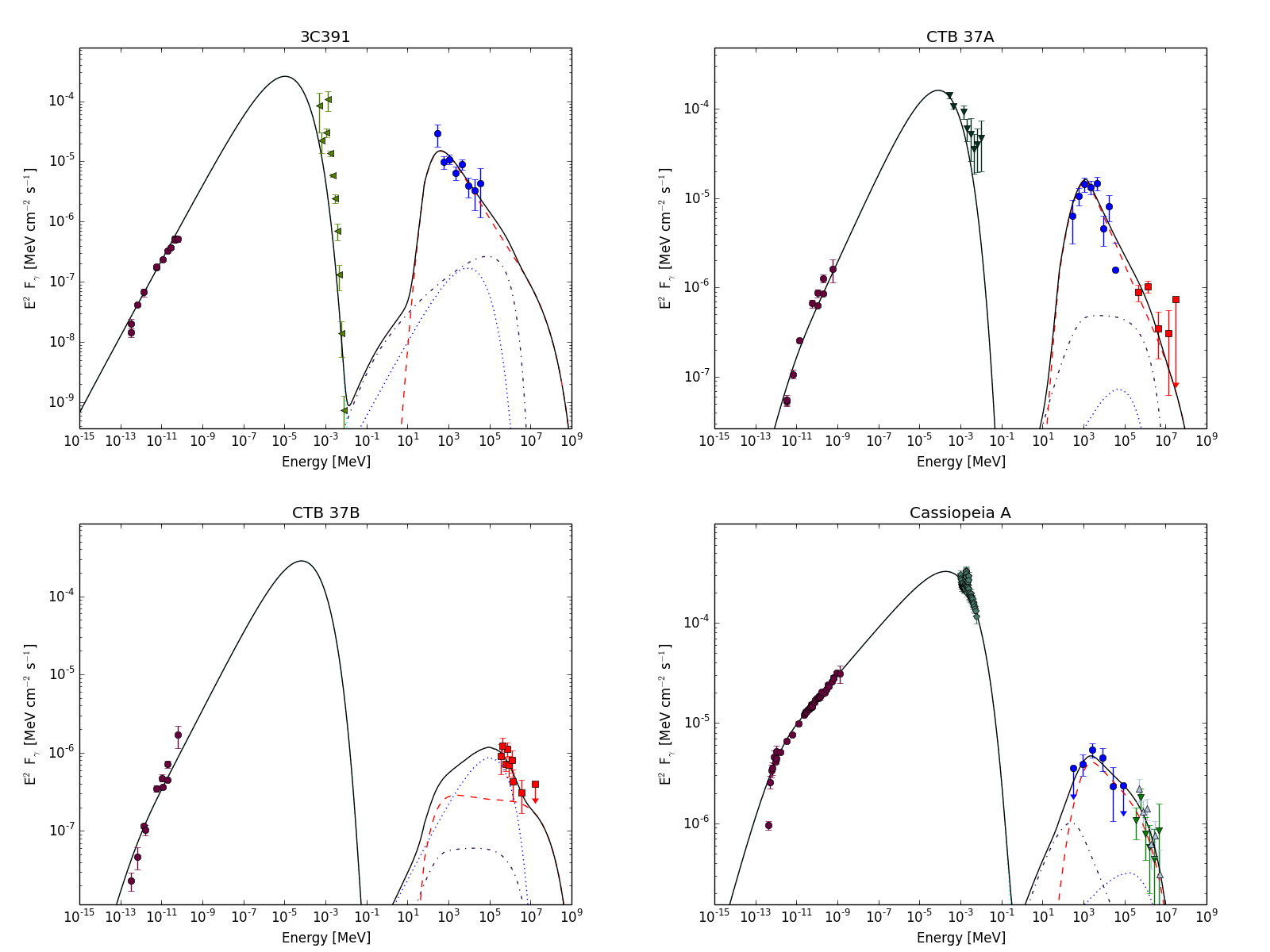}
 \end{center}%
 \caption{SEDs for the sources
   3C391, CTB37A, CTB37B and Cas A, corresponding to the the hadronically
   dominated case as indicated via
   a vertical line in
   Fig.\ \ref{etot0:fig}. The black
   line shows the total emission,
   fully made up by synchrotron
   radiation for the low-energy
   bump of the SED and composed of
   $\pi^{0}-$ decay photons (red,
   dashed line), bremsstrahlung
   (black, dot-dashed line) and
   Inverse Compton scattering
   (blue, dotted line). References for the fitted data points
   are given in Table \ref{electrons:tab}. Data are indicated as - 
   radio range: brown filled bullets; X-ray range: sideways-to-left-pointing,
   olive triangles (ASCA), sideways-to-right-pointing
   green triangles (XMM), downward-pointing, dark green
   triangles (Suzaku), upward-pointing, bright green triangles
   (ROSAT),  steel blue diamonds (Chandra); high-energy range: blue,
   filled bullets (Fermi), orange diamonds (AGILE), red squares (H.E.S.S.), downward-pointing, green
   triangles (MAGIC), upward-pointing, light-blue
   triangles (VERITAS) and pink stars (Milagro). \label{sed0:fig}}
\end{figure}%

\begin{figure}[ht!]%
 \begin{center}%
 \includegraphics[width=\linewidth]{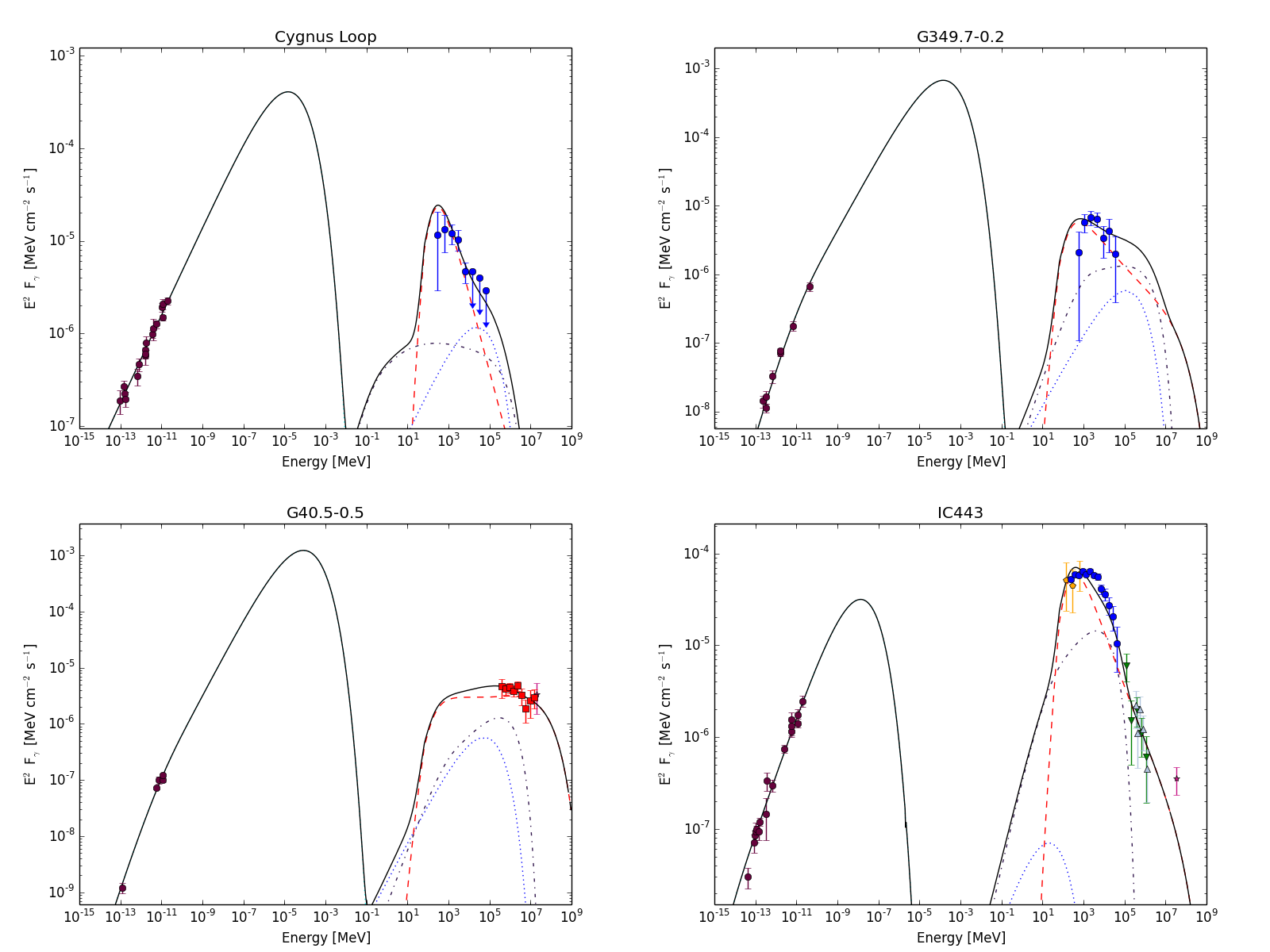}
 \end{center}%
 \caption{SEDs for the sources Cygnus Loop, G349+0.2, G40.5-0.5 and IC443, corresponding to the
   the hadronically   dominated case
   as indicated via  a vertical line
   in
   Fig.\ \ref{etot1:fig}. Labeling
   as in Fig.\ \ref{sed0:fig}.  \label{sed1:fig}}
\end{figure}%
\begin{figure}[ht!]%
 \begin{center}%
 \includegraphics[width=\linewidth]{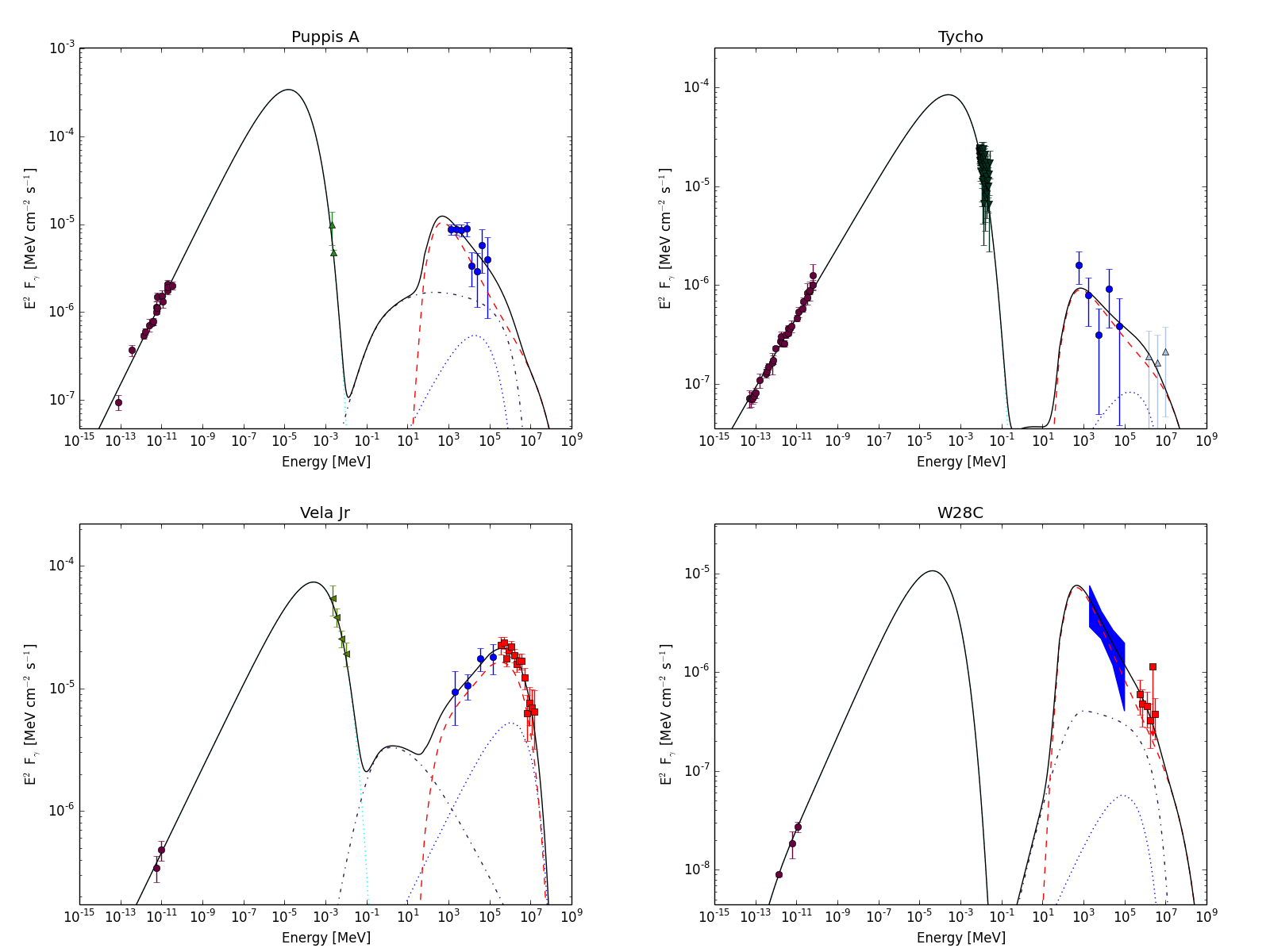}
 \end{center}%
 \caption{SEDs for the sources Puppis A, Tycho, Vela Junior and W28C, corresponding to the
   the hadronically   dominated case
   as indicated via  a vertical line
   in
   Fig.\ \ref{etot2:fig}. Labeling
   as in Fig.\ \ref{sed0:fig}.  \label{sed2:fig}}
\end{figure}%
\begin{figure}[ht!]%
 \begin{center}%
 \includegraphics[width=\linewidth]{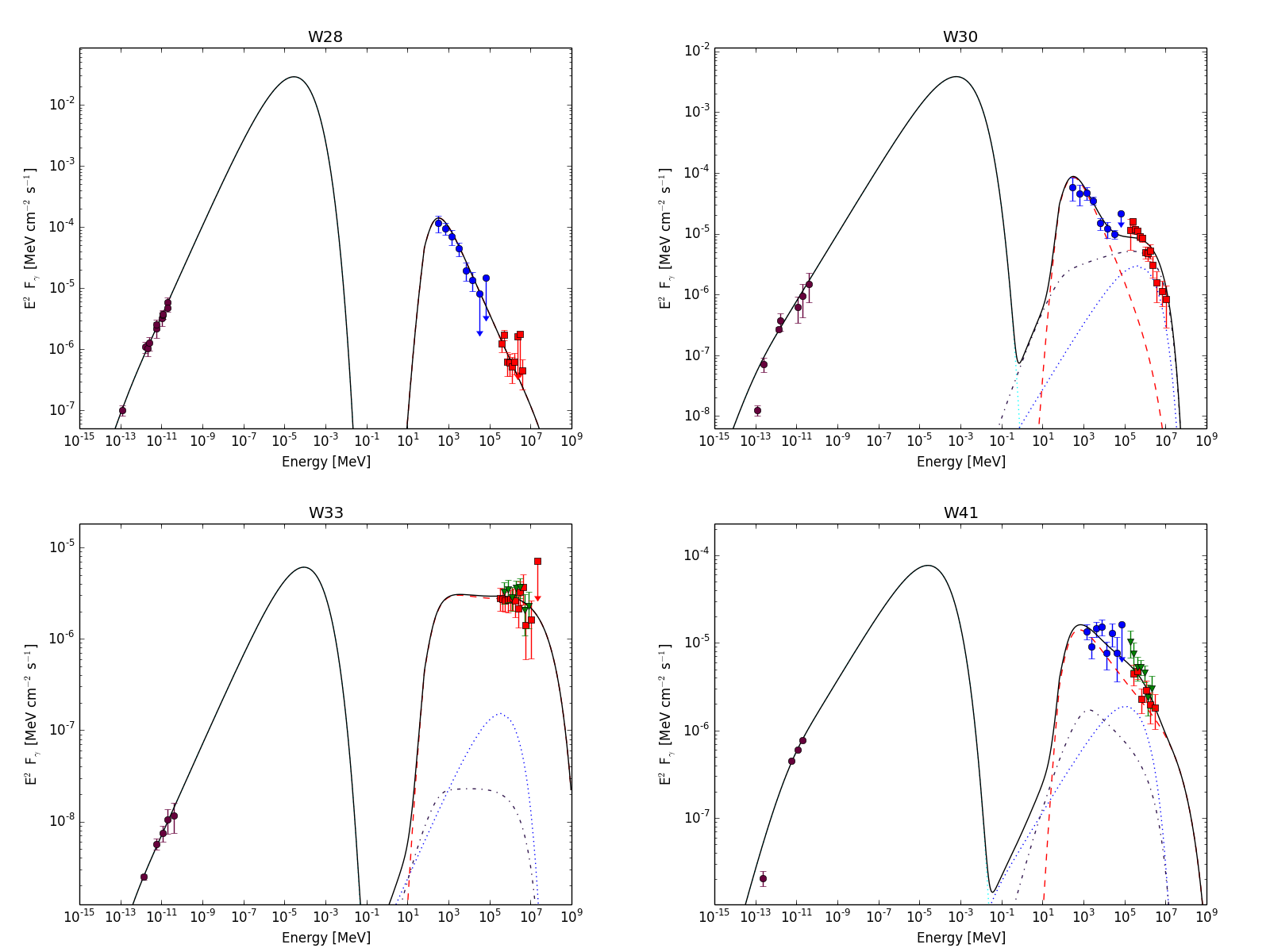}
 \end{center}%
 \caption{SEDs for the sources W28, W30, W33 and W41, corresponding to the
   the hadronically   dominated case
   as indicated via  a vertical line
   in
   Fig.\ \ref{etot3:fig}. Labeling
   as in Fig.\ \ref{sed0:fig}.  \label{sed3:fig}}
\end{figure}%

\begin{figure}[ht!]%
 \begin{center}%
 \includegraphics[width=\linewidth]{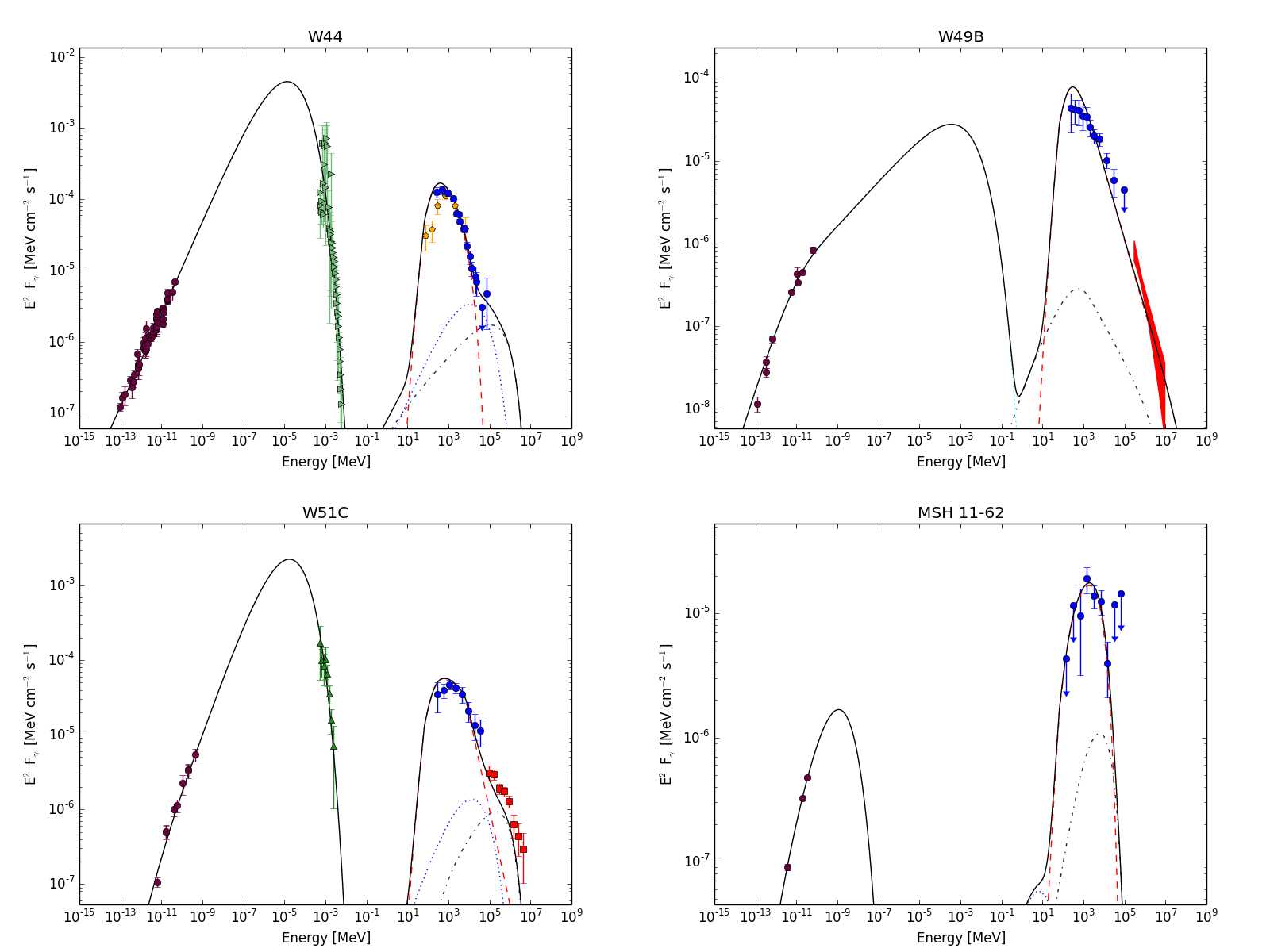}
 \end{center}%
 \caption{SEDs for the sources W44, W49B, W51C and MSH 11-62, corresponding to the
   the hadronically   dominated case
   as indicated via  a vertical line
   in
   Fig.\ \ref{etot0:fig}. An exception is W44 (see above), where $B=120$~$\mu$Gauss was chosen. Labeling
   as in Fig.\ \ref{sed0:fig}.  \label{sed4:fig}}
\end{figure}%

\begin{figure}[ht]%
 \begin{center}%
 \includegraphics[width=\linewidth]{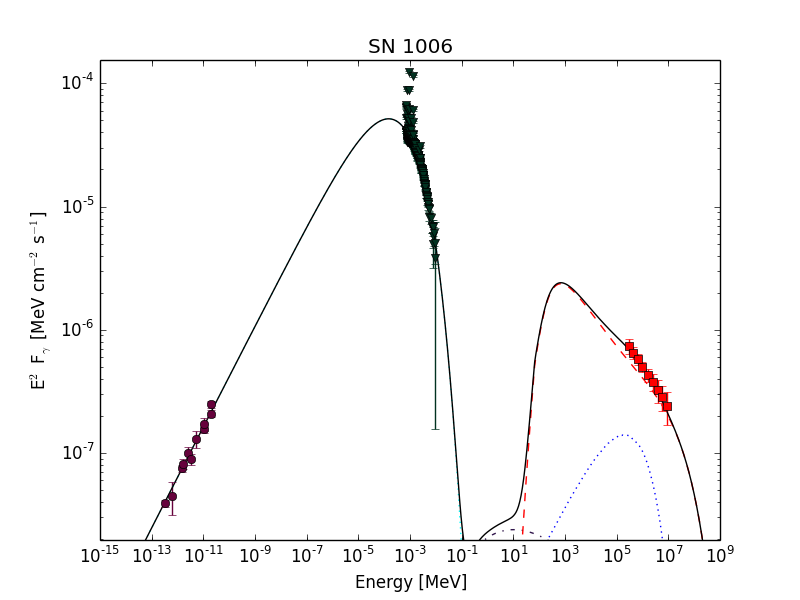}
 \end{center}%
 \caption{SEDs for the source SN1006, corresponding to the
   the hadronically   dominated case
   as indicated via  a vertical line
   in
   Fig.\ \ref{etot_sn1006:fig}. Labeling
   as in Fig.\ \ref{sed0:fig}.  \label{sed_sn1006:fig}}
\end{figure}%
\clearpage

\bibliography{lib,tmp_julia}

\end{document}